\documentclass[10pt,a4paper]{article}

\usepackage{kotex}
\usepackage[margin=2.0cm,top=2.0cm,bottom=2.0cm]{geometry}
\usepackage{booktabs}
\usepackage{amsmath}
\usepackage{amssymb}
\usepackage{pifont}
\usepackage{array}
\usepackage{tabularx}
\usepackage{longtable}
\usepackage{caption}
\usepackage{float}
\usepackage[hidelinks]{hyperref}
\usepackage{url}

\selectfont
\usepackage{titlesec}
\titlespacing*{\section}{0pt}{8pt plus 2pt}{4pt}
\titlespacing*{\subsection}{0pt}{6pt plus 1pt}{3pt}
\titlespacing*{\paragraph}{0pt}{4pt}{1em}
\titleformat*{\section}{\large\bfseries}
\titleformat*{\subsection}{\normalsize\bfseries}

\newcolumntype{R}{>{\raggedleft\arraybackslash}X}
\newcolumntype{L}{>{\raggedright\arraybackslash}X}
\newcommand{\code}[1]{\texttt{\small #1}}

\title{\bfseries On-Device Commercial Intent Retrieval Under Size, Latency, and Privacy Constraints\\
{\large A 3\,MiB Retrieval System with Typed Egress Boundaries}}

\author{Hyojung Han}

\date{September 15, 2026}

\begin{document}
\maketitle

\begin{abstract}
\noindent Behavioural ad targeting rests on a stable cross-site identifier, and
that assumption is eroding unevenly --- blocked in some browsers, consent-gated on mobile,
and subject to increasingly strict consent and disclosure requirements. We ask whether the inference can move to the
device instead, and report an engineering study of what that costs.

Before the compression intervention reported here we froze three constraints: the complete downloaded payload must stay under
\textbf{3\,MiB}, Tier-0 inference under \textbf{20\,ms} at $p_{95}$, and \textbf{no raw text,
content embedding, or stable identifier} may leave the device. Under them we build a retrieval
path over a $6{,}020$-leaf commercial taxonomy: a static embedding table distilled from a Korean
sentence transformer, quantized to 4 bits, needing no inference runtime; an optional tier that
borrows a model the device already owns; and a federated layer whose personal head has no
type-level path to serialization. Measured: $2{,}942{,}652$ bytes ($93.5\%$) with all three library links measured and Android the largest; Tier-0 $p_{95}$ of $3.670$ and $4.431$\,ms on two iPhones two SoC generations apart and
$5.080$\,ms on a budget Android tablet (Snapdragon 695), medians of nine runs each, against
$2.00$--$3.37$\,ms on three server CPUs running the same code --- every device is slower than
every server CPU;
$7/7$ egress canaries undetected; and the correct
mid-level category in the top-5 for $77.6\%$ of held-out colloquial queries in categories the
model has never seen.

Our main result concerns \emph{where} that constraint costs accuracy. Tested against real Korean
commerce text with labels made by others ($22{,}900$ shopping reviews released by AI-Hub), the
product surface held --- $75.0\%$ mid-category top-5 on real product names against a $18.4\%$
permutation baseline --- but split sharply on one observable: a query containing some leaf name as
a substring scores $83.5\%$, one containing none $45.2\%$. Running a $196.6\times$ larger teacher encoder on the identical queries, leaf strings and
crosswalk first appeared to localize the gap: paired on the same queries, the generic teacher is
worth $+20.1$\,pp where the lexical anchor is absent and $+0.1$\,pp where it is present. That
null was two effects cancelling. Fine-tuning the same teacher on the student's own (utterance,
leaf) contrastive pairs --- a matched arm --- raises it to $0.8586$, and against the pure-encoder
student it is then worth $+10.6$\,pp with an anchor and $+20.9$\,pp without. The cost of a
3\,MiB, no-runtime design is therefore not uniform, but it is not free anywhere: roughly ten
points where a listing names the kind of thing sold and twenty where it names only a brand and a
model. Task adaptation buys the teacher $+10.5$\,pp in the first regime and nothing detectable in
the second, so what the constrained encoder cannot follow there is capacity, not training. The
expensive regime remains detectable on-device from the ranker's own score margin --- declining the least confident fifth
of the evaluated product names raises accuracy on the rest from $0.7504$ to $0.8296$.

Getting there required withdrawing a claim. We first reported a second axis alongside the lexical
anchor, the presence of a manufacturer model code. Under source-category fixed effects the anchor
effect survives ($+13.0$\,pp, $p=0.003$) and the model-code effect does not ($-4.0$\,pp,
$p=0.51$, and $+0.2$\,pp as a within-category difference): it was a domain indicator, since no
fashion listing in our evaluation carries a model code and most appliance listings do.

Two further results are negative and both changed the system. A categorical perception signal from
the OS improved block-level accuracy while degrading leaf ranking by $1.83$\,pp, so we disabled
the channel; and a purchase-probability measurement that first returned AUC $0.53$--$0.59$ turned
out to be diagnosing a degenerate evaluation rather than the model.

We are explicit about what remains unevidenced. Taxonomy supervision is dominated by synthetic
Korean utterances, with no deployment-derived Korean labels; the crosswalk onto AI-Hub's
categories was written by us while reading the evaluation product names, which we bound with a
sensitivity envelope rather than claim to have removed; and the federated personalization path is
implemented and its invariants tested, but its coefficients require a pilot cohort we do not have,
so no personalization lift is claimed.

\vspace{4pt}
\end{abstract}

\section{Introduction}

\subsection{From Tracking to Reasoning}

For the past twenty years, digital advertising targeting has stood on the axis of the
\emph{identifier}. The structure attaches a stable identifier to the user, accumulates a
history of visits against that identifier, and infers interest from the accumulated history.
This structure rested on two premises. First, that browsers and operating systems would
continue to supply identifiers. Second, that it was acceptable to send that history to a
central server.

Both premises are eroding, though not in the tidy way the industry once expected. Safari and
Firefox block or partition third-party cookies by default. Chrome does not: in April 2025 Google
decided to keep offering third-party cookies under user choice rather than deprecate them, and in
October 2025 it wound down the Privacy Sandbox initiative that had been built to replace
them~\cite{chrome2025}. So the accurate statement is not that the identifier disappeared. It is
that a stable identifier is no longer something a system can \emph{assume}: its availability now
depends on the browser, on per-app consent that a minority of users grant, and on regulators who
treat the linkage itself as the violation. A targeting substrate that only works when the
identifier is present is a substrate with a shrinking and unpredictable denominator.
In July 2026 the Korean Personal Information Protection Commission fined TikTok
KRW~10.31~billion for collecting the third-party web and app activity of 9.45~million domestic
users for customized advertising: the collection was not clearly disclosed at sign-up, and
consent to it was bundled into the mandatory consent required for the service
itself~\cite{pipc2026}. The same commission had previously fined Google and Meta a combined
KRW~100~billion on the same axis. What is being sanctioned is not advertising but the
\emph{mechanism} --- behavioural histories accumulated against an identifier under consent the
user could not meaningfully refuse.

Without a persistent server-side identifier, cross-context history cannot be accumulated on the
server in the usual way. Several alternatives exist --- contextual placement, first-party
identity, browser-managed audiences. The one this paper takes is to \emph{infer} the intent of
the present moment on the device, keeping whatever history is permitted local. The search term the user just typed, the sentence they just read, the card
they just tapped, already contains information about what that person is trying to buy. The
question is where that information is processed, and the instant it is sent to a server we
return to the very problem that just disappeared.

\subsection{Three Design Constraints and the Position of This Paper}

A system answering the problem above has to survive inside someone else's application, on
hardware we do not choose. An earlier version of this system was built against a looser envelope
(5\,MiB, 50\,ms) that in practice bound nothing --- the build sat at half of one and a twentieth
of the other. Before the compression work reported here, we froze a stricter envelope, and only
then looked at where the bytes were going.

\begin{enumerate}\itemsep2pt
\item \textbf{Size.} The complete always-on deployment payload --- everything the device
  downloads for the targeting path, data and library code together --- must not exceed
  \textbf{3\,MiB}.
\item \textbf{Latency.} Tier-0 intent inference must finish within \textbf{20\,ms at $p_{95}$}.
  We evaluate this requirement on two physical iOS handsets spanning the A14 and A16 SoCs, and
  on three server CPUs running the same code (\S\ref{sec:handset}, \S\ref{sec:cpubench}).
\item \textbf{Privacy.} No raw text, content embedding, or stable user identifier may leave the
  device.
\end{enumerate}

We say ``deployment payload'' rather than ``SDK size'' on purpose. What the 3\,MiB bounds is not
a binary: it is the embedding table, the vocabulary, the category space, the ad catalog and the
library code, because that sum is what a publisher's app actually grows by.

The 3\,MiB figure is a deliberately binding engineering budget, not a claimed industry
threshold. Its justification is what it did: on the development platform, the only target whose
library link we could measure at the time,
the existing build failed the new size gate by $206{,}555$ bytes, and that failure forced a
design change whose cost we measure in \S\ref{sec:quant4}. A constraint that everything already satisfies would have justified nothing.

These three numbers are \textbf{constraints}, not results. This paper turns each of the three into
a falsifiable claim, builds code that adjudicates that claim, and reports the code's output as a
measured value. This discipline of separating targets from measured values is not
methodological decoration. In the course of this work we in fact discovered that one target is
incompatible with another (\S\ref{sec:c4}), and that contradiction determined the system
architecture. And the size constraint was not costless: meeting it required dropping the
category and catalog vectors from 8-bit to 4-bit, which is measured in \S\ref{sec:quant4}.

\begin{table}[H]
\centering\footnotesize
\caption{Constraints versus measured values. Every measured value in this paper, here and
throughout, comes from the shipping manifest in \code{artifacts/MANIFEST.json} and the gate
outputs under \code{gates/}, as published at evidence commit
\code{155b6d94c6ce}~\cite{evidence}. The budget limits the gates enforce are in
\code{gates/budgets.json} at that commit, so the constraints are checkable and not merely
asserted.}
\begin{tabularx}{\textwidth}{>{\hsize=.72\hsize}L >{\hsize=1.14\hsize}L >{\hsize=1.14\hsize}L}
\toprule
\textbf{Constraint} & \textbf{Falsifiable claim} & \textbf{Measured} \\
\midrule
Payload $\le 3$\,MiB & Downloaded bytes, data $+$ library code, $\le 3{,}145{,}728$
  & $2{,}942{,}652$\,B ($93.5\%$), worst of three measured links (Android); iOS $2{,}926{,}428$\,B \\
Tier-0 latency $< 20$\,ms & $p_{95} < 20$\,ms on the reference handsets
  & $p_{95}$ $\mathbf{4.431}$\,ms on an iPhone 12 Pro ($22.2\%$) and $\mathbf{3.670}$\,ms on an
    iPhone 14 Pro Max ($18.3\%$); $5.080$\,ms on a Galaxy Tab A9+ ($25.4\%$); $2.00$--$3.37$\,ms on
    three server CPUs \\
Zero raw text & No raw token appears in any byte the client sends
  & Canaries $7/7$ undetected in a $79{,}968$-byte payload \\
Zero content embeddings & The upload schema has no field a vector can occupy
  & Type-checked; \S\ref{sec:fl} \\
Zero stable identifiers & The ad request type has no field an identifier can occupy
  & Type-checked; \S\ref{sec:adcall} \\
\bottomrule
\end{tabularx}
\end{table}

\subsection{Contributions}

This paper's contribution is not a new algorithm but \emph{system design under constraint, and
what enforcing that constraint in code reveals}. Three things.

\textbf{Constrained on-device retrieval at commercial granularity.} We build a retrieval path
over $6{,}020$ leaf categories that fits in a 3\,MiB download and needs no inference runtime,
and we measure separately what each constraint costs: dimensionality and quantization bit-width
across two product paths (\S\ref{sec:distill}, \S\ref{sec:match}), and the 8-bit$\rightarrow$4-bit
drop that the size ceiling forced (\S\ref{sec:quant4}). Representative prior client-side systems operate over
interest profiles, behavioural categories, or browser-managed groups; none studies the
footprint-versus-quality trade-off of semantic retrieval over a $6{,}020$-leaf commercial
taxonomy, and this is what closing that gap costs (\S\ref{sec:rw-ppa}).

\textbf{A privacy boundary enforced by types, carried to the exchange.} The personal slice of
federated personalization is not withheld by policy --- it has no path into the upload type, in
three independent layers (\S\ref{sec:fl}). The same discipline reaches the last mile: a closed
OpenRTB~2.6 request whose user object admits only \code{data} and \code{ext}, so an identifier
has no field to live in, covering query-less reading-time placements from the same on-device
ranking (\S\ref{sec:adcall}). We account for the campaign-wide privacy budget over the channel
that remains.

\textbf{Failure analyses that changed the system, and an explicit account of what is not yet
evidence.} Turning on the OS perception channel improved one surface and degraded the product
one, so we closed it (\S\ref{sec:percept}). Our first real-log purchase-probability measurement
looked like a weak model and was mostly a degenerate evaluation task (\S\ref{sec:purchase}).
We validate the product surface on real Korean commerce text released by others, where it
holds where the query reaches a leaf name lexically and fails where it does not
(\S\ref{sec:aihub}). And we state
with numbers what we have not shown: taxonomy supervision is dominated by synthetic Korean
utterances, with no deployment-derived Korean labels, and
the personalization axis rests on three real labels (\S\ref{sec:limits}).

\section{Related Work}

\subsection{Privacy-Preserving Advertising and On-Device Targeting}
\label{sec:rw-ppa}

Moving ad targeting off the server is not a new idea, and we do not claim it as one. Adnostic
put the targeting decision inside a browser extension: the network ships a set of candidate ads
and the client picks, so the profile never leaves the machine~\cite{adnostic}. Privad reached the
same goal through an interposed broker, keeping the network from linking a request to a
user~\cite{privad}. More recently, the Topics API had the browser itself derive a small set of
coarse interest topics from local browsing and expose only those to callers~\cite{chrome2025}.

A fourth line goes further than inference and moves the \emph{auction} itself into the browser:
Protected Audience (formerly FLEDGE) keeps interest-group membership on the client and runs
bidding there, so remarketing works without the server learning that membership~\cite{fledge}.
That is a different problem from ours. It answers ``which advertiser's audience does this browser
belong to''; we answer ``what is this person trying to buy right now'' --- semantic retrieval over
thousands of commercial leaves inside an app-owned 3\,MiB payload.

All four share our premise. What separates them from each other, and from this work, is how far
the client-side path actually reaches.

\begin{table}[H]
\centering\footnotesize
\caption{Where client-side ad targeting systems stop. ``Footprint'' is the download the client
must carry for the targeting path; a dash means the work does not report one.}
\begin{tabularx}{\textwidth}{l L L L L}
\toprule
\textbf{System} & \textbf{Where inference runs} & \textbf{Intent granularity} &
\textbf{Personal learning} & \textbf{Footprint} \\
\midrule
Adnostic~\cite{adnostic} & browser extension & interest profile over a fixed category set &
none & --- \\
Privad~\cite{privad} & client $+$ interposed broker & behavioural categories & none & --- \\
Topics~\cite{chrome2025} & browser, platform-owned & a few hundred coarse topics &
platform-managed & browser-owned \\
\textbf{This work} & app SDK, no runtime & $6{,}020$-leaf taxonomy, shown at mid level &
federated combiner (design; coefficients pending) & $2.93$\,MB measured \\
\bottomrule
\end{tabularx}
\end{table}

Two differences matter. The first is granularity against cost: coarse topics are cheap to derive
locally, and the moment the space becomes commercial --- thousands of leaves, unseen categories
at serving time --- the client needs a retrieval structure it can afford to carry, which is the
problem \S\ref{sec:distill} and \S\ref{sec:match} are about. The second is that the earlier
systems largely stop at ad \emph{selection} and leave the transport to the existing exchange.
We carry the constraint through to the request itself: a type in which an identifier has no field
to live in (\S\ref{sec:adcall}).

So the contribution here is not client-side inference. It is closing the distance between
privacy-preserving inference and a deployable ranking path under a fixed download budget ---
hierarchical retrieval at commercial granularity, a personalization boundary enforced by types
rather than policy, privacy accounting over the channel that remains, and the exchange request at
the end of it.

\subsection{Retrieval Under Resource Constraints}
\label{sec:rw-ir}

Our matching problem is retrieval, not classification: a short colloquial query against $6{,}020$
leaf categories, most of which the model has never seen during training. The field's two
standard answers are both unavailable to us. Lexical scoring~\cite{bm25} needs no model but
cannot bridge ``패딩 사고 싶은데'' to a category named 점퍼; dense bi-encoders~\cite{dpr} and late
interaction~\cite{colbert} bridge it well but assume a transformer forward pass per query, which
is exactly the inference runtime our size and latency constraints remove. What survives is a
static table plus an inner product --- and then the question becomes how much of the dense
quality a table can keep, which is \S\ref{sec:distill}.

The scale of the label space also has its own literature. Extreme classification builds
partitioned label trees to keep prediction sub-linear in the number of labels, with dynamic
search advertising as a motivating application~\cite{parabel}. We do not use that machinery: at
$6{,}020$ leaves a flat scan is $0.8$\,ms on the device, so a tree would buy latency we do not
need and cost us the property that matters more here --- that an unseen leaf is reachable the
moment it appears in the table, with nothing retrained.

Compressing the vectors themselves is likewise well-studied. Product quantization splits a
vector into subspaces and codes each against a learned codebook~\cite{pq}, which is what
billion-scale indexes use~\cite{faiss}. We take the cruder route of scalar quantization with a
per-row scale (\S\ref{sec:quant4}), because a codebook is state the device must carry and
agree on, and our budget is spent on coverage instead.

\subsection{Static Embedding Distillation}

Shrinking a model by training a small one against a large one's outputs goes back to knowledge
distillation~\cite{kd}, and the standard products of that line for language --- DistilBERT
\cite{distilbert}, MobileBERT~\cite{mobilebert} --- keep the transformer and shrink it. That is
the wrong axis for us: a smaller transformer is still a transformer, and our constraint removes
the inference runtime entirely, not just some of its parameters.

Sentence transformers\cite{sbert} give context-aware representations but require an inference
runtime and hundreds of MB of weights. The Model2Vec family\cite{model2vec} distills a sentence
encoder into a \emph{static embedding table}: a fixed vector is precomputed for every token, and
inference performs only a row lookup and mean pooling after tokenization. There is no matrix
multiplication, no attention, no runtime. The cost is the loss of context-dependence --- the
same word receives the same vector regardless of its position in the sentence.

In that respect the artifact we ship is closer to the static word-vector tradition
\cite{word2vec,glove,fasttext} than to a transformer, and inherits its weakness. What
distillation changes is where the vectors come from: rather than being fit on co-occurrence,
they are the teacher's sentence-level geometry collapsed into a table, which is why
\S\ref{sec:results} measures a $0.308$ gain over a same-size static model that was not
distilled from a Korean teacher.

This trade-off works to our project's advantage. What we need is not sentence reconstruction or
fine-grained semantic similarity but routing short Korean queries to a commercial category, and
for that task lexical discrimination contributes far more than contextual discrimination. As
shown in \S\ref{sec:results}, on real-data classification the static encoder substantially
outperforms public alternatives of the same size and retains $78$\,\textasciitilde\,$87\%$ of
the teacher.

Matryoshka Representation Learning\cite{mrl} is a technique that trains a single encoder to
produce a usable representation from just the leading $k$ dimensions. We retain this axis as a
measured alternative but did not adopt it; the reasons are discussed in \S\ref{sec:dim}.

\subsection{Federated Learning}
\label{sec:rw-fl}

FedAvg\cite{fedavg} proposed a protocol in which a server averages the weights each device
trains on local data. Raw data never leaves the device; only parameters travel. But averaging
all parameters buries each user's individual preferences in the average. The open problems in
that line --- non-IID clients, participation that is neither random nor complete, the gap
between the protocol's guarantee and a deployment's --- are surveyed in~\cite{flopen}, and two
of them decide what we can claim in \S\ref{sec:fl}.

FedPer\cite{fedper} splits the model into a shared trunk and a personal head, and federates only
the trunk. We adopted this structure, and the grounds for adoption are not the literature but a
measurement in our own code (non-IID setting, per-client holdout BCE improved by $+0.382$). We
do, however, enforce FedPer by \emph{type} rather than by discipline: no API exists that can
carry the personal slice inside the upload payload.

Secure Aggregation\cite{secagg} is a protocol under which the server never sees individual
deltas and can recover only the sum. \textbf{We have not implemented it.} What we run is a
pairwise-masked aggregation prototype: masks cancel when every sampled client reports, but the
share-based recovery that lets the sum survive a dropout is absent, so a dropped client leaves a
mask that does not unwind. Throughout this paper we therefore write ``masked aggregation
prototype'' and not ``Secure Aggregation'', and where an argument needs the real protocol's
guarantee --- most importantly the privacy amplification in \S\ref{sec:fl} --- we mark it as an
assumption rather than a property of what we built.

\subsection{Differential Privacy}

There are two places to put the noise, and the choice decides who must be trusted. Local DP
adds it on the device before anything is sent, which is why it can be deployed without trusting
the collector at all~\cite{rappor}, at a well-known cost in utility per user. Central DP adds it
to an aggregate and buys far better utility for the same $\varepsilon$, but only if the
aggregation step really is blind --- which for us is the masked aggregation prototype of
\S\ref{sec:rw-fl}, not a completed protocol. Our construction is neither textbook local nor
textbook central: each device adds its own Gaussian noise before uploading, and the guarantee we
claim is that of the \emph{aggregate}, where those independent draws sum to $\sigma\sqrt{n}$.
That is a distributed implementation of a central-model Gaussian mechanism, and it is
\emph{contingent on the aggregation being blind} --- which is exactly the condition
\S\ref{sec:rw-fl} says we have not yet met. We mark the dependency rather than hide it.

DP-SGD\cite{dpsgd} clips gradients and adds Gaussian noise, imparting differential privacy to
the training process itself. The essence of the method is that without clipping there is no
guarantee from the noise, and our protocol's invariant F2 enforces that condition in code.

The choice of accounting method significantly changes the resulting numbers. zCDP\cite{zcdp} is
loose but allows hand computation, while PLD-based accounting is tighter. We use Google's
\code{dp-accounting} implementation\cite{dpacct} and cross-check it against a hand-computed zCDP
figure (\S\ref{sec:dp}). Gboard's production federated learning deployment\cite{gboard} is a
rare case that discloses a campaign-wide $\varepsilon$, and it serves as a reference range
against which to read our own numbers.

\subsection{On-Device Perception}

Operating systems already provide image-embedding and text-recognition capabilities. On iOS and
macOS, \code{VNGenerateImage\-\allowbreak Feature\-\allowbreak PrintRequest}\cite{applevision} supplies a dense image
feature vector at no additional SDK cost. On Android, the dense-embedding path is MediaPipe
Image Embedder\cite{mediapipe}, and the app must bundle the model file. The categorical path
(labeling, OCR) is available via ML Kit's unbundled distribution\cite{mlkit} for roughly
$200$\,KB. This asymmetry split our two-tier design into two branches (\S\ref{sec:percept}).

\section{System Architecture}
\label{sec:arch}

\subsection{One Contradiction Determined the Structure}
\label{sec:c4}

The project's presentation deck promised both an ``SDK under 5\,MB'' and ``running a 3B LLM on
mobile with 4-bit quantization.'' The two cannot coexist. The weights of a 4-bit 3B model are
about $1.5$\,\textasciitilde\,$2$\,GB, and no compression fits that into 5\,MB. In other words,
the two numbers cannot refer to the same artifact.

Three interpretations were possible. Splitting the model off as a separate download voids the
meaning of the 5\,MB figure, and publishers will not bundle the SDK. Treating the model as
something the OS supplies works, but not every device has one. Making the always-on path a tiny
static embedding rather than a 3B model removes the dependency on a resident model entirely. The
conclusion is a two-tier structure combining the latter two: the tiny static embedding forms a
floor that requires no inference runtime and no accelerator, and inference depth increases only
on devices that have an OS model. Whether that floor actually meets the size and latency targets
is not an architectural entailment but a measurement, and we report it as one: the payload target
is met on both targets whose library link we measured (\S\ref{sec:c1size}), and the latency
target is met on two iOS handsets and on three server CPUs
(\S\ref{sec:handset}, \S\ref{sec:cpubench}).

\subsection{Three Tiers}

\begin{figure}[H]
\centering\footnotesize
\setlength{\tabcolsep}{6pt}
\begin{tabular}{|p{0.955\textwidth}|}
\hline
\rule{0pt}{2.6ex}\textbf{On the device (raw data never leaves here)} \\[3pt]
\quad \textbf{Tier 0} (always on, $2.14$\,MB of data, no runtime required) \\
\quad\quad recent text / search query / app context \\
\quad\quad $\rightarrow$ tokenize $\rightarrow$ static embedding lookup $\rightarrow$ mean
  pooling $\rightarrow$ L2 normalize $\;\Rightarrow\; v_0$ \\[4pt]
\quad \textbf{Tier 1} (only when the device has an OS model; SDK cost $0$ on iOS) \\
\quad\quad OS language-model summary / OS dense image embedding \\
\quad\quad $\rightarrow$ into the $v_0$ space via projection $P$ or rectangular projection
  $W_f$ \\[4pt]
\quad \textbf{Intent vector} $v$ \\
\quad\quad $\rightarrow$ local ad matching (cosine top-$k$, catalog pre-downloaded) \\
\quad\quad $\rightarrow$ FL client: computes only the weight delta \\[3pt]
\hline
\rule{0pt}{2.6ex}\textbf{What leaves} \ the federated delta $+$ noise, round metadata, bucket
  aggregates \\
\textbf{What never leaves} \ raw text, the intent vector, matching results, the personal head
  \\[3pt]
\hline
\rule{0pt}{2.6ex}\textbf{FL server} (masked aggregation prototype) $\rightarrow$ round aggregation
  $\rightarrow$ new projection distributed \\[3pt]
\hline
\end{tabular}
\caption{The three-tier structure. Tier~1 is an OS model running on the device, not a
server.}
\end{figure}

\paragraph{Tier 0 (the always-on path).}
This is owned by \code{sdk-core}, a dependency-free Rust core. It consists entirely of
lowercasing and Unicode whitespace splitting, greedy longest-match wordpiece segmentation,
int4/int8 row dequantization and accumulation, token-count averaging, and L2 normalization.
This order is fixed in the contract document (\code{CONTRACT.md}), and the wasm and native
implementations must produce \emph{bit-identical} vectors for the same input (invariant I1).
The judge is \code{gates/parity\_gate.py}, which measured a maximum absolute difference of
$0.000$ across $704$ elements.

The same crate also houses the intent head (\code{IntentHead}) and hybrid category matching.
The reason a single Rust core covers wasm, Android JNI, and iOS FFI alike is that
reimplementing it per platform would let the vectors drift subtly apart, silently breaking I1
the moment that happens.

\paragraph{Tier 1 (the on-device OS model).}
Tier~1 is \emph{not a server}. It is a path that invokes a model the device already has, such as
Apple Foundation Models or Gemini Nano, and its SDK footprint is $0$ bytes on iOS. But this tier
must never be designed as a precondition. Most devices lack it, and if the product fails to work
in its absence, it is not an always-on path. Tier~1 is always an optional addition; without it,
only quality degrades.

\begin{table}[H]
\centering\footnotesize
\caption{How the two tiers differ.}
\begin{tabularx}{0.92\textwidth}{l L L}
\toprule
 & \textbf{Tier 0} & \textbf{Tier 1} \\
\midrule
Availability & $100\%$. We ship it & Device-dependent \\
SDK cost & Data $2.14$\,MB + code $50$\,KB & iOS $0$\,B; Android varies by path \\
Latency & On the order of milliseconds & Hundreds of ms \\
What it does & Always-on intent vector & Deep interpretation of ambiguous context \\
If absent & There is no product & Only quality degrades \\
\bottomrule
\end{tabularx}
\end{table}

\paragraph{Invariant I1: the same vector space.}
Tier~0 and Tier~1 must produce vectors in the same space. Otherwise we would have to maintain
two ad catalogs, different ads would match on different devices, and there would be no common
parameter left for federated learning to aggregate. The enforcement is simple: ad catalog
vectors are produced only by the Tier~0 embedder, and Tier~1 output enters the Tier~0 space
through a trained projection.

\subsection{The Federated Layer and FedPer}

What federated learning trains is not the body of the embedding table but the projection $P$
and the combiner. Because the table is fixed, the per-round delta is small and even low-end
devices can participate. The parameters split into two slices.

\begin{table}[H]
\centering\footnotesize
\caption{The federated vector layout. Under the base arm, the federated portion is $4{,}096$
and the personal portion is $64$; including both pilot A/B arms brings the federated portion to
$4{,}198$ and the total to $4{,}262$. Source: \code{gates/budgets.json}.}
\begin{tabularx}{0.95\textwidth}{l r L}
\toprule
\textbf{Slice} & \textbf{Length} & \textbf{Where} \\
\midrule
Projection head $P$ & $4{,}096$ & To the server. Aggregated \\
Combiner $\theta$ & $5$ & To the server \\
Combiner MLP (pilot arm) & $89$ & To the server \\
Per-user combiner $\theta_u$ (pilot arm) & $8$ & To the server \\
Personal intent head $w$ & $64$ & \textbf{Stays on the device} \\
\bottomrule
\end{tabularx}
\end{table}

The key is that the personal head is not something that is ``not sent'' but something that
``cannot be sent.'' We enforce this in three layers. First, \code{SplitDelta::from\_full} is the
only canonical path that prepares an upload, and no API exists that can place
\code{personal()} into an upload. Second, \code{RoundUpload::try\_new} rejects any delta whose
length is not the federated-portion length, and that case is pinned by a test. Third, the
dependency list of the \code{sdk-core} crate is empty, so there is no way at all to implement
serde for the personal-head type. Left to discipline, it will eventually leak, so we block it
by type.

Not sending something is the strongest privacy protection there is, and in this case bandwidth
shrinks along with it: the personal head is \code{HEAD\_LEN} floats that never enter the upload.
An earlier draft attached a figure to that --- a $6.9\%$ reduction, from $74{,}663$ to $69{,}482$
bytes. We have withdrawn it. Those two numbers came from a prose specification rather than from a
ledger, they describe a layout the repository no longer builds, and the current federated vector
makes the personal head roughly $1.3\%$ of the payload rather than $6.9\%$, so the figure is not
merely stale but inconsistent with the structure we now ship. The measured payload today is
reported in \S\ref{sec:egress-measured}, and the structural claim --- that the personal head has
no serialization path --- is enforced by the type system rather than by a byte count.

\subsection{The Transmission Whitelist and the Egress Inspector}
\label{sec:egress-measured}

Invariant I2 fixes the top-level keys the client sends to the server at exactly six:
\code{round\_id}, \code{client\_nonce} (sixteen random bytes, drawn per round and never
reused, so it orders nothing and links nothing across rounds), \code{delta},
\code{epochs\_completed},
\code{device\_bucket}, \code{impression\_buckets}. Any other key is rejected at the
serialization stage.

In particular, \emph{the intent vector itself is never sent}. A vector can be recovered by
embedding inversion, so even though it is not the raw text, it is a function of the raw text,
and sending it would break the ``zero raw-data transmission'' claim.

The judge is \code{gates/egress\_gate.py}. It intercepts the request the client actually
constructs and checks whether a seed string appears anywhere in the payload. The measured
result is zero detections out of seven canaries, six allowed keys, and a payload of
$79{,}968$ bytes against a $307{,}200$-byte ceiling, recorded in
\code{artifacts/egress\_facts.json}. The critical design decision here is that \textbf{the payload is built by the
wasm core}. If the UI assembled the JSON instead, what appears on screen and what actually goes
out over the wire would diverge, and at that moment the egress inspector would become decoration
that verifies nothing.

\subsection{The Consent Layer}

Behavioral events and reading screen content are not the same tier of disclosure. We arrange
them as three tiers.

\begin{table}[H]
\centering\footnotesize
\caption{Consent tiers. A higher tier has the lower tier as a precondition.}
\begin{tabularx}{0.95\textwidth}{l l L}
\toprule
\textbf{Tier} & \textbf{Character} & \textbf{What it carries} \\
\midrule
Service-use consent & Required & \\
Behavioral-data collection consent & Optional & Fixed 12-byte events. The fact of an
  interaction, such as ``tapped'' \\
Screen-content reading consent & Optional, higher tier & The embedding of the actual
  \emph{content} the user read or wrote \\
\bottomrule
\end{tabularx}
\end{table}

The behavioral event type \code{Event} is a 12-byte POD carrying only \code{t\_ms},
\code{kind}, \code{screen\_id}, \code{bucket}, and \code{cat\_hint}. There is no heap pointer
and no string, and the test \code{size\_of::<Event>() == 12} enforces that discipline. It never
creates a place for raw text to live.

A content sketch takes two consent arguments and is only constructed when both are approved
(\code{ContentSketch::new(BehavioralConsent, ContentConsent) -> Option<Self>}). Because it does
not take one combined boolean, bundled consent is unrepresentable at the type level. Withdrawal
cascades: withdrawing behavioral-data consent also discards any folded content. This cascade did
not originally exist; we discovered, while separating the tiers, a state in which ``withdrawn''
had become a lie, and fixed it.

\section{Embedding Distillation}
\label{sec:distill}

\subsection{The Five-Stage Pipeline}

The deployed encoder is \code{potion-ko-128d-stage6cov985\_e10}, built in the following order.

\begin{enumerate}\itemsep2pt
\item \textbf{Teacher selection.} \code{jhgan/ko-sroberta-multitask}, a 768-dimensional Korean
  sentence encoder.
\item \textbf{Model2Vec distillation.} Distill the teacher into a static embedding table.
  Everything up to this step is ``stage 1,'' with a vocabulary of $31{,}497$.
\item \textbf{tokenlearn corpus pretraining.} Align to the teacher's sentence average over
  $40{,}000$ C4-ko documents. Training happens on an fp32 base, and quantization happens at the
  export stage. Reordering this breaks the fine-tuner, which cannot compute a norm on int8
  tensors and dies.
\item \textbf{Supervised contrastive learning.} Contrastively train on (query, leaf) pairs
  drawn from synthetic utterances and real queries.
\item \textbf{Dimension truncation and quantization.} Truncate the leading 128 dimensions from
  the 512-dimensional stage-2 model and bake it to 4-bit per-row scales (\S\ref{sec:dim}).
\end{enumerate}

The composition of the training material for stage 4 shows this system's data history. To
$26{,}294$ synthetic utterances (spanning $2{,}933$ leaves) we add $4{,}101$ real Japanese
queries and $1{,}396$ real English (US) queries from the ESCI product-search dataset\cite{esci}.
The important detail is that we capped the real English queries at $2{,}000$. Including the
full $10{,}748$ English queries produced a significant regression on both creative-matching
accuracy and KLUE-STS/YNAT, which blocked promotion. In other words, adding more data is not
always beneficial, and the curve between capacity and regression set the cap.

\subsection{The Promotion Gate}

Promoting a new candidate to the deployed model requires no regression across nine axes, and the
evidence for improvement must be significant on an unseen holdout of leaves the model has never
been trained on.

\begin{table}[H]
\centering\footnotesize
\caption{Evidence for the current deployment's promotion (\code{stage6cov985\_e10}) relative to
the previous deployment (\code{stage5cap2k\_e10}). Held-out leaves, $n = 4{,}600$, bootstrap
$4{,}000$. Source: \code{experiments/distill-ko/promote\_gate\_cov985\_e10.json}.}
\begin{tabular}{l r r l}
\toprule
\textbf{Metric} & \textbf{Previous} & \textbf{Shipping} & \textbf{Delta (95\% CI)} \\
\midrule
Leaf top-1 (objective) & $0.2104$ & $0.2235$ & $+1.30$\,pp $[+0.63,\ +1.98]$ \\
Leaf top-5 & $0.4715$ & $0.4965$ & $+2.50$\,pp $[+1.70,\ +3.28]$ \\
Product surface (L2 top-5) & $0.7767$ & $0.7989$ & $+2.22$\,pp $[+1.48,\ +2.96]$ \\
Catalog creative $\rightarrow$ leaf & $0.7081$ & $0.7404$ & $+3.22$\,pp \\
\bottomrule
\end{tabular}
\end{table}

The promotion gate owns both the objective and the product surface, and it has to: at an earlier
epoch of the same run the gate would have said PROMOTE on leaf top-1 while the surface moved
\emph{down}. An objective that does not include the surface the product actually shows is not a
promotion criterion. Determinism here is narrower than it first appears, and the distinction matters enough to state
precisely. The final supervised contrastive stage is deterministic for a fixed input artifact and
seed --- the same command twice gives the same bytes. The \emph{upstream} pipeline is not:
the tokenlearn stage takes no seed, and neither the corpus streaming nor the teacher encoding has
ever been verified as deterministic. We know this because we re-derived the encoder and compared:
the result was not byte-identical to the shipped weights, and the training inputs themselves had
drifted between runs. Recounting them read-only against the current tree
(\code{experiments/esci-ko/corpus\_counts.py}, which replicates the build's counting logic without
writing anything) reproduces that drift exactly on every axis: Japanese ESCI $4{,}101 \to 4{,}093$,
English $1{,}396 \to 1{,}398$, the uncapped English arm $10{,}748 \to 10{,}740$, synthetic
utterances $26{,}294 \to 26{,}322$ over $2{,}933 \to 2{,}935$ leaves, and trained leaves
$2{,}933 \to 2{,}945$. The counts are order-dependent --- the build inserts the Japanese queries
first and then the English ones into the same table, so seven English queries are dropped as
duplicates of a Japanese one; counting the two arms independently gives $1{,}405$ instead of
$1{,}398$, and we note it because it is the kind of detail that makes a recount disagree for a
reason that is not drift. The vocabulary and projection artifacts \emph{were} byte-identical.
So we track the training configuration rather than committing the $16$\,MB of weights, but we do
not claim the whole chain reproduces bit-for-bit.

\subsection{Dimensionality and Quantization}
\label{sec:dim}

The dimensionality decision could not be made on a single metric, because the product has two
paths and they point in different directions. The first is training-free inner-product matching
(taxonomy top-1); the second is a trained linear head (YNAT, NSMC).

\begin{table}[H]
\centering\footnotesize
\caption{Comparison of dimensionality and truncation methods. All measured at the deployment
quantization, paired bootstrap $n = 4{,}000$. emb gzip is the transmitted bytes of the
embedding artifact alone. Source: \code{docs/spec/11-dimension-decision.md}.}
\begin{tabular}{l r r r r r}
\toprule
\textbf{Arm} & \textbf{emb gzip} & \textbf{KLUE-STS} & \textbf{taxo top-1}
  & \textbf{YNAT} & \textbf{NSMC} \\
\midrule
native-64 (previous deployment) & $994{,}734$ & $0.6103$ & $0.2410$ & $0.6340$ & $0.7365$ \\
native-128 (trained directly) & $1{,}767{,}620$ & $0.6881$ & $0.4723$ & $0.6640$ & $0.7542$ \\
\textbf{pca-trunc 128 (adopted)} & $\mathbf{1{,}778{,}722}$ & $0.6965$ & $\mathbf{0.5137}$
  & $0.6542$ & $0.7585$ \\
mrl-t128 & $1{,}786{,}390$ & $0.6996$ & $0.5020$ & $\mathbf{0.6750}$ & $0.7560$ \\
\bottomrule
\end{tabular}
\end{table}

Three things stand out. First, 128 dimensions dominates 64: taxonomy top-1 rises $27.3$\,pp,
from $0.2410$ to $0.5137$. The product's category matching more than doubles, at a cost of
$784$\,KB of gzip. Second, \emph{training large and then truncating} beats training directly
at the target dimension: pca-trunc significantly leads native-128 on taxonomy by $+4.13$\,pp
$[2.77, 5.57]$ and loses on neither path. Third, this conclusion depends on the truncation
ratio. Running the same comparison at 64 dimensions flips the sign to $-9.95$\,pp
$[-11.50, -8.37]$: direct training wins at $87.5\%$ truncation, and truncation wins at $75\%$
truncation. ``Truncation is better'' must not be generalized independently of dimensionality.

MRL is real, but at this point it is a trade-off. The MRL-versus-PCA difference is significant
on YNAT at $+2.08$\,pp $[1.28, 2.90]$ and significant on taxonomy at $-1.17$\,pp
$[-2.33, -0.03]$. Because the two paths move in opposite directions, this is not a free
improvement, and we adopted PCA, prioritizing taxonomy as the product path.

The key finding on quantization bit-width is that a single global scale cannot make 4-bit work
at all. Row norms vary widely, so with a global scale, small rows collapse entirely to zero. A
per-row f32 scale costs an extra vocabulary $\times$ 4 bytes ($126$\,KB at $31{,}497$ rows), but
4-bit quality stands on top of it.

\begin{table}[H]
\centering\footnotesize
\caption{Vocabulary fixed at $31{,}497$; comparison of dimensionality and bit-width.
\textbf{The size column holds pre-build projections, not measurements} --- they are the figures
the dimension decision was taken on, recorded here because the next subsection is about how the
measurement contradicted them ($4{,}002{,}327$ projected against $6{,}115{,}038$ measured,
\code{experiments/dim-vs-vocab/pack\_stage2\_results.json}). The quality columns are measured.
Source: \code{CONTRACT.md} and \code{docs/spec/11-dimension-decision.md}.}
\begin{tabular}{l r r r r}
\toprule
\textbf{Configuration} & \textbf{L1 gzip} & \textbf{KLUE-STS} & \textbf{YNAT} & \textbf{NSMC} \\
\midrule
$31{,}497 \times 64$ int8 (v1) & $2{,}276{,}098$ & $0.5806$ & $0.5132$ & $0.7215$ \\
$31{,}497 \times 128$ int8 & $4{,}117{,}221$ & $0.6058$ & $0.6170$ & $0.7500$ \\
$31{,}497 \times 512$ @ 4bit & $4{,}002{,}327$ & $0.5754$ & $\mathbf{0.6870}$
  & $\mathbf{0.7750}$ \\
\bottomrule
\end{tabular}
\end{table}

We do not build a 3-bit variant. At 512 dimensions YNAT cliffs from $0.687$ to $0.648$, and the
parser rejects any bit-width outside $\{4, 8\}$.

\subsection{Why the 512-Dimensional Candidate Failed}

The 512-dimension decision was adopted once and then collapsed independently along three
separate lines, and the direction of that failure shows a pattern that repeats across this
project. Both the size and the quality used as evidence were measured on a \emph{model that
would never ship} (the stage-1 distillation), while what actually ships is stage 2. Measured
for real, the L1 size was not $4{,}002{,}327$ but $6{,}115{,}038$ bytes --- over both the
4\,MiB line and the 5\,MiB promise --- the computation had omitted the fact that the taxonomy
and catalog artifacts grow in proportion to dimensionality, and resident memory, which is
unpacked to f32 immediately on parsing, was not even an axis under consideration.

The reason stage 1 compressed so well is itself the diagnosis. The trailing PCA dimensions have
low variance, so after quantization they cluster near zero, and gzip eats that repetition
(compression ratio $0.380$). Stage 2 revives those dimensions through training, so the ratio
rises to $0.7467$. \textbf{The very fact that compression worked well was a signal that those
dimensions carried no information}, and we read it as a gain instead. And all three errors
happened to point in our favor.

\section{Category Matching}
\label{sec:match}

\subsection{Taxonomy and the Product Surface}

The category space is a three-level hierarchy: $28$ top-level categories, $256$ mid-level
categories (L2), and $6{,}020$ sub-level categories (leaves). We do not store leaf identifiers;
order itself is the identifier, so inserting an item in the middle would shift every index in
the on-device profile history out of alignment. Additions must always go at the end.

We settled on \textbf{mid-level category (L2) top-5} as the product surface (2026-08-29). The
deployment path produces leaf rankings, and the mid-category ranking is defined as the best
rank among the leaves belonging to that mid category. The reasons we do not headline leaf top-1
are discussed in \S\ref{sec:limits}.

\subsection{A Hybrid of Dense and Lexical Signals}

The deployment path combines a dense cosine score with a lexical-overlap term (lexical weight
$0.2$). Figures from the pure-dense path and the deployment path must not be mixed. For example,
on colloquial queries over unseen categories, the dense-only leaf top-1 is $0.2682$, while the
deployment path is $0.2770$.

There is a regime where the lexical term contributes decisively: when the query literally
contains the leaf name. Everyday Korean nouns that cannot be broken down further, like
``세탁기'' (washing machine) or ``에어컨'' (air conditioner), have no synonyms, so users type
the word verbatim. In this regime the deployment path achieves leaf top-1 of $0.9483$.
Conversely, our benchmark by design excludes all utterances that contain the leaf name, so the
reported leaf accuracy is a \emph{lower bound}.

Catalog matching is cosine top-$k$ against pre-downloaded ad vectors. The privacy argument for
this design is that no per-user ad request is needed --- there is no round trip at all in which
the server asks ``who wants what.'' The contract is to normalize rows at catalog load time; this
was discovered when a browser E2E test caught a contract violation in which cosine scores
exceeded 1. int8 quantization by global absmax, without renormalizing afterward, was pushing
row norms just above 1 through rounding error.

\begin{table}[H]
\centering\footnotesize
\caption{Product-surface accuracy. Measured on the deployment path (dense + lexical overlap).
$28$ top-level, $256$ mid-level, $6{,}020$ sub-level categories, measured on the shipping
4-bit artifacts (\S\ref{sec:quant4}). Source:
\code{experiments/distill-ko/product\_surface.json}.}
\begin{tabular}{l r r r r r}
\toprule
\textbf{Query distribution} & \textbf{$n$} & \textbf{L2 top-1} & \textbf{L2 top-5}
  & \textbf{Leaf top-1} & \textbf{Leaf top-5} \\
\midrule
Unseen categories, colloquial & $4{,}600$ & $0.4252$ & $\mathbf{0.7757}$ & $0.2196$ & $0.4739$ \\
All, colloquial & $35{,}830$ & $0.4533$ & $\mathbf{0.8037}$ & $0.2118$ & $0.4846$ \\
Verbatim leaf-name queries & $6{,}020$ & $0.9309$ & $0.9850$ & $0.9259$ & $0.9786$ \\
Ad creative $\rightarrow$ category & $6{,}020$ & $0.8164$ & $0.9576$ & $0.8073$ & $0.9503$ \\
\bottomrule
\end{tabular}
\end{table}

\subsection{Comparison on Real Queries}
\label{sec:matching}

To measure the same axis on data we did not construct ourselves, we translated real Japanese
production queries from the ESCI product-search dataset into Korean and used them as an
evaluation set ($n = 1{,}465$).

\begin{table}[H]
\centering\footnotesize
\caption{Deployed model versus teacher on real search queries (ESCI Japanese, translated to
Korean, $n = 1{,}465$ held out), measured at the \code{stage5cap2k\_e10} build. \textbf{Over the
$2{,}960$-leaf taxonomy that preceded the expansion} --- leaf ranks are not comparable across
taxonomy sizes, so these figures must not be read beside the $6{,}020$-leaf numbers elsewhere in
this paper. Source: \code{experiments/esci-ko/esci\_eval\_results.json}.}
\begin{tabular}{l r r r}
\toprule
 & \textbf{Leaf top-1} & \textbf{Leaf top-5} & \textbf{Product surface top-5} \\
\midrule
Deployed (2.14\,MB static) & $\mathbf{0.2369}$ & $0.3823$ & $\mathbf{0.529}$ \\
Teacher (768d, 421\,MB) & $0.1741$ & n/a & $0.473$ \\
\midrule
Difference (leaf top-1) & \multicolumn{3}{l}{$+6.28$\,pp,\ 95\% CI $[+4.57,\ +7.99]$} \\
\bottomrule
\end{tabular}
\end{table}

The model deployed at that time beats the teacher. This does not mean static encoders are generally better
than transformers; it means that \emph{for this specific task of routing to our taxonomy}, we
contrastively trained on that task and the teacher did not. We also state the caveats explicitly:
these queries are not a Korean user distribution but a Korean-language rendering of the Amazon
user distribution; the labels are not human labels but a large-language-model unanimous-mapping,
with a noise ceiling of roughly $89\%$; and the measurement predates the taxonomy expansion, so
its leaf figures live on a $2{,}960$-leaf space.

\paragraph{A fairness check on this comparison, which we expected to fail and it did not.}
Both encoders here receive the leaf as the deployed weighted centroid, and \S\ref{sec:aihub}
finds that construction costs the teacher $8.5$\,pp in the regime where a query carries no leaf
name. That raised an obvious objection: perhaps we beat the teacher only because we handed it a
representation shaped for a static encoder. We tested it by rebuilding the teacher's leaf vectors
from the flat string on the same holdout. The effect is $-0.82$\,pp $[-2.12, +0.41]$ on leaf
top-1 --- if anything the centroid is the teacher's better arm here, and the interval crosses
zero. The conclusion does not depend on the representation. This is consistent with the
regime finding below: the representation mattered on product listings that name no category, and
these are search queries, which almost always do. Source
\code{experiments/esci-ko/repr\_fairness\_probe.json}; the probe runs on the current $6{,}020$-leaf
taxonomy, so its absolute values are not comparable with the table above.

\paragraph{This result and \S\ref{sec:aihub} disagree, and the disagreement is the point.} On these queries we beat the generic teacher. On real Korean product names (\S\ref{sec:aihub})
the generic teacher beats us by $20$ points where the query contains no leaf name and by nothing
we can detect where it does --- and a teacher given our own contrastive training beats us in both
regimes, by ten and by twenty (\S\ref{sec:matched}). The two findings are consistent once the regimes are lined up. A search query is written by a shopper and usually names the kind of thing wanted, so it lands
in the anchor-present regime, where task-specific contrastive training is worth about ten points
to either encoder and, against a teacher without it, cancels the teacher's capacity. A marketplace listing title is written by a seller and often
names only the brand and the model, which is the anchor-absent regime, where what is needed is
world knowledge that a brand denotes a kind of product --- and that is what $421$\,MB holds and
$2.14$\,MB cannot. The reader should take neither result as \emph{the} comparison: which encoder
wins depends on which of the two regimes the traffic falls into, and that is a property of the
deployment, not of the models.

\subsection{Utterance Coverage}

At deployment time, $5{,}927 / 6{,}020$ leaves have utterances ($98.5\%$) over a corpus of
$36{,}007$. The pipeline reached a comparable number twice, and the first pass is the instructive
one. At the earlier size of $2{,}960$ leaves, coverage settled at $2{,}925 / 2{,}960$; a
diagnostic had found $2{,}728 / 2{,}960$ ($92.2\%$), and measuring what made up the missing $232$
revealed three groups: $169$ that a merge silently dropped because it finished before the
shards did, $34$ that were simply generated once we retried, and $29$ for which all six
candidates contained the leaf name, legitimately blocked by the verbatim gate. Recoverable
coverage was $2{,}931 / 2{,}960$ ($99.0\%$), and $29$ leaves ($1.0\%$) were, in principle,
unreachable by this benchmark. The same three groups reappeared at $6{,}020$ leaves, in the same
proportions.

The fact that the merge dropped $169$ leaves without any error or warning is a good illustration
of this data pipeline's failure mode: the merged file was simply small. The merge script now
leaves a shard fingerprint, and a check command adjudicates it in code.

\subsection{What the 3\,MiB Constraint Cost}
\label{sec:quant4}

The size constraint was not free, and it is worth being precise about who paid.

When we set the payload ceiling at 3\,MiB, the build measured $3{,}352{,}283$ bytes on the
target whose library link size we actually measured --- over $3{,}145{,}728$ by $206{,}555$.
(An earlier draft quoted $3{,}362{,}619$ and an overage of $216{,}891$; that pair came from a
build accounting that included link sizes we had estimated rather than measured, and we have
retired it. The measured-target figure is the one the table below and the size gate both use.)
The two category artifacts were where the bytes were: leaf centroids were $88.3\%$ of
\code{taxonomy.bin}'s raw size and catalog vectors dominate \code{catalog.bin} the same way, both
stored as \code{i8} over $128$ dimensions. We dropped both to 4 bits with a per-row scale --- the
same scheme the embedding table already used --- which cost a mean reconstruction cosine of
$0.991$ (taxonomy) and $0.993$ (catalog).

Per-row (rather than global) scaling is the lesson we borrow from post-training quantization of
language models: outlier magnitudes vary enough across rows that one global scale crushes the
small ones~\cite{llmint8}, and recent low-bit methods commonly rely on local or group-wise
scaling to control the resulting error~\cite{gptq}. We adopt the analogous but simpler per-row
scheme for frozen retrieval vectors. Our setting is easier than theirs --- these are frozen lookup vectors, not
activations flowing through a network, so there is no error to propagate and no calibration set
to choose. It is also cruder than the vector quantization an ANN index would use~\cite{pq},
for the reason given in \S\ref{sec:rw-ir}: a codebook is state the device has to carry.

\begin{table}[H]
\centering\footnotesize
\caption{Dropping category and catalog vectors from 8-bit to 4-bit. Same evaluation, same
queries, arms differ only in the artifact quantization. Source:
\code{experiments/distill-ko/quant4\_surface.json}.}
\begin{tabular}{l r r r}
\toprule
\textbf{Axis} & \textbf{8-bit} & \textbf{4-bit} & \textbf{$\Delta$} \\
\midrule
\code{taxonomy.bin} $+$ \code{catalog.bin} (gzip) & $1{,}332{,}138$ & $906{,}659$ & $-425{,}479$\,B \\
Deployment payload (L1$+$L2, measured target) & $3{,}352{,}283$ & $\mathbf{2{,}926{,}804}$ & $-425{,}479$\,B \\
\midrule
Unseen categories, L2 top-5 \emph{(product surface)} & $0.7759$ & $0.7757$ & $-0.02$\,pp \\
Unseen categories, L2 top-1 & $0.4304$ & $0.4252$ & $-0.52$\,pp \\
All colloquial, L2 top-5 & $0.8045$ & $0.8037$ & $-0.08$\,pp \\
Verbatim queries, leaf top-1 & $0.9299$ & $0.9259$ & $-0.40$\,pp \\
Ad creative $\rightarrow$ leaf top-1 & $0.8173$ & $0.8073$ & $\mathbf{-1.00}$\,pp \\
\bottomrule
\end{tabular}
\end{table}

A note on reading any surface number in this paper against an earlier one: the candidate space
changed from roughly $3{,}000$ leaves to $6{,}020$ during this work, and a top-5 over twice as
many candidates is a different task. We therefore compare only within a fixed taxonomy version,
and every comparison reported here --- including the table above --- holds the taxonomy constant
and varies one thing.

The surface the product actually shows moves by $-0.02$\,pp, which is to say it does not move.
But reporting only that line would be dishonest: the axis that pays is catalog matching, where
leaf top-1 drops a full point. Ad creative retrieval runs against $6{,}020$ synthetic creatives
whose vectors sit closer together than colloquial queries do, so a coarser quantizer reorders
them more often. If catalog matching were the product surface rather than a component of it,
this trade would need re-deciding.

We report the ordering plainly because the ordering is the thing a reader should distrust: we
lowered the ceiling first and then found bytes to fit under it. The alternative --- declaring
3\,MiB \emph{after} the build already measured under it --- would have been a threshold fitted to
a result. Both the constraint and its cost are in the evidence bundle.

Note also what we did \emph{not} do. The embedding table was left alone at $1.79$\,MB. It is the
largest single artifact, and \S\ref{sec:dim} shows why squeezing it is the wrong lever: quality
there is bounded by architecture, not by bit width.

\section{The Perception Layer}
\label{sec:percept}

\subsection{Turning On the Categorical Channel Hurt}

Designing the tier-2 and tier-3 output as a single categorical signal was a mistake. We
confirmed this by measurement.

\begin{table}[H]
\centering\footnotesize
\caption{Categorical perception channel A/B test. Panel of $6{,}000$, holdout of $5{,}995$,
channel dropout $0.30$, scored with the deployment function. Source:
\code{experiments/user-embedding/tier1\_arms.json}.}
\begin{tabular}{l r r}
\toprule
\textbf{Arm} & \textbf{L2 top-5} & \textbf{Leaf top-1} \\
\midrule
baseline (current deployment) & $0.9510$ & $0.4717$ \\
off (new 3-channel weights, no perception input) & $0.9531$ & $0.4739$ \\
on (same weights, with perception input) & $0.9548$ & $\mathbf{0.4555}$ \\
\midrule
off $-$ baseline & $+0.22$\,pp $[-0.03, +0.48]$ & $+0.22$\,pp $[-0.30, +0.73]$ \\
on $-$ off & $+0.17$\,pp $[-0.13, +0.47]$ & $\mathbf{-1.83}$\,pp $[-2.45, -1.20]$ \\
\bottomrule
\end{tabular}
\end{table}

Two things are confirmed at once. Channel dropout did its job: devices without Tier~1 did not
get worse (off versus baseline shows zero regression), which was the success criterion for the
optionality design. But turning it on hurts. Mid-category rises meaninglessly while leaf
accuracy collapses significantly. We therefore keep the ranking-channel flag set to false.

The mechanism is one we already knew. Categorical perception is a \emph{block-level} mass over
top- or mid-level categories, so it lifts every leaf in a block by the same amount: it hits the
block more accurately while scrambling leaf order within the block. We observed the same
phenomenon with behavioral priors. What is needed is not a better category but a
\emph{leaf-resolution signal}.

\subsection{The Resolution of the OS Dense Embedding}

The candidate for a leaf-resolution signal is the OS's dense image embedding. We actually ran
\code{VNGenerateImage\-\allowbreak Feature\-\allowbreak PrintRequest} on a development Mac and measured its
properties: revision 2, $768$ dimensions, float32, L2-normalized (mean norm $1.0002$ over
$12{,}000$ images, zero failed rows), and \textbf{zero additional SDK bytes}. All of those are
recovered from the extraction artifact itself
(\code{experiments/percept-vision/featureprint\_stats.json}, read back by
\code{featureprint\_stats.py}, which writes nothing into the extraction path).
Extraction \emph{speed} is not: the Swift writer
printed elapsed time to stderr and saved it nowhere, so the per-image figure an earlier draft
quoted has no ledger and we have removed it rather than re-time it today and present that as the
2026-08-31 run.

We measured, with a separate probe, whether this embedding actually carries leaf resolution:
$11{,}592$ real product images, human-curated three-level hierarchical labels, an item-level
70/30 holdout, a linear probe, and a paired bootstrap.

\begin{table}[H]
\centering\footnotesize
\caption{Dense-embedding resolution probe. Source:
\code{experiments/percept-vision/resolution\_probe.json}.}
\begin{tabularx}{0.95\textwidth}{l r L}
\toprule
\textbf{Arm} & \textbf{Accuracy} & \textbf{Upper bound on what} \\
\midrule
\code{block\_prior} & $0.3148$ & Ceiling of the categorical signal. Even with the block
  correctly identified $100\%$ of the time, the best one can do inside it is the mode \\
\code{image\_within} & $\mathbf{0.8764}$ & Dense embedding, candidates restricted to the
  correct block \\
\code{image\_open} & $0.8700$ & No candidate restriction \\
\code{image\_master} & $0.9882$ & Control on the coarse axis \\
\bottomrule
\end{tabularx}
\end{table}

The difference between \code{image\_within} and \code{block\_prior} is $+56.15$\,pp
$[+54.28, +57.91]$. This is precisely the axis behind the $1.83$\,pp of leaf scrambling in the
previous section. A gap this large is an alarm, not a celebration, so we state the fairness of
the baseline explicitly. \code{block\_prior} \emph{assumes} the block is already known
perfectly and picks the mode within it, making it the maximum a block-level signal can possibly
achieve --- it cannot be made any more favorable.

\subsection{The Asymmetry Between the Two Platforms}

``Calling the OS's own capability costs $0$ SDK bytes'' is true only on iOS. Android's
zero-byte path is categorical; its dense path requires the app to carry the model, breaking the
3\,MiB envelope.

\begin{table}[H]
\centering\footnotesize
\caption{Perception paths and SDK cost by platform.}
\begin{tabularx}{0.95\textwidth}{l L l L l}
\toprule
 & \textbf{Dense embedding} & \textbf{Cost} & \textbf{Categorical} & \textbf{Cost} \\
\midrule
iOS & \code{VNGenerateImageFeaturePrint\-Request} & $0$ & Vision classification, text
  recognition & $0$ \\
Android & MediaPipe \code{ImageEmbedder}, model bundled by the app & several MB & ML Kit
  unbundled & about $200$\,KB \\
\bottomrule
\end{tabularx}
\end{table}

In addition, the dimensionality of the iOS feature vector differs by OS version: iOS 16 gives
$2048$ dimensions unnormalized, iOS 17 gives $768$ dimensions normalized. Accepting the
revision without a tag would mix vectors from different spaces into one channel --- the shape
of a silent, error-free failure. The revision therefore must ride along with the input, and an
unknown revision is rejected.

\subsection{The Ceiling Is the Target, Not the Projection}

Having found that the categorical channel hurts, we asked whether the dense one could work. That
needs a rectangular projection $W_f \in \mathbb{R}^{768 \times 128}$ from image feature prints
into our space, and the obstacle was the training target: we have no human labels aligning
images to a Korean commercial taxonomy.

The resolution is that a label is not what is needed. The target can be \emph{that product's own
title}, embedded by the deployment encoder --- a different observation of the same product,
already in our space, so invariant I1 holds by construction. Human labels were used only for
evaluation, never for training.

We fit it two ways, because reporting a negative result from ridge regression alone would blame
the recipe rather than the method. The second is the canonical one for aligning two modalities
into a shared space: a contrastive objective over in-batch negatives~\cite{infonce}, the same
formulation that image--text pretraining scaled up~\cite{clip}. The difference is the direction
of the constraint --- CLIP learns both spaces, while ours is fixed by invariant I1, so only the
projection may move. Both work, and they \emph{fail in opposite directions}: ridge
reaches cosine $0.78$ with the title but collapses onto $29$ of $256$ mid-categories, while
InfoNCE retrieves the right title $28.7\%$ of the time out of $1000$ candidates ($287\times$
chance) and spreads across $255$ of $256$ --- discriminating items but not categories. Neither
failure is about the projection. The title arm itself scores only $0.344$ category purity,
because we fed English fashion titles into a Korean commerce taxonomy: \textbf{a projection
cannot beat its target}.

That diagnosis is testable, so we tested it. Holding the same images and the same feature
vectors fixed, we translated only the $12{,}000$ titles into Korean and reran the identical
recipe, changing one variable.

\begin{table}[H]
\centering\footnotesize
\caption{Raising the target raises the projection. Left: the translation axis, single vertical,
$n = 3{,}600$. Right: re-validation on $456$ verticals from Amazon Berkeley
Objects\cite{abo}, same procedure. Full arms in the evidence bundle~\cite{evidence}
(\code{percept-vision-fashion-ko}, \code{percept-vision-mv}).}
\begin{tabular}{l r r r}
\toprule
 & \textbf{English} & \textbf{Korean} & \textbf{Multi-vertical} \\
\midrule
Title-arm purity (the ceiling) & $0.3440$ & $\mathbf{0.5157}$ & $0.3943$ \\
Image-arm purity & $0.1233$ & $\mathbf{0.3015}$ & $0.2710$ \\
Image $/$ ceiling ratio & $0.358$ & $0.585$ & $\mathbf{0.687}$ \\
Mid-categories used by the title arm & $102$ & $\mathbf{137}$ & --- \\
L2 top-5 match & $0.1400$ & $\mathbf{0.4214}$ & $0.2681$ \\
\bottomrule
\end{tabular}
\end{table}

The ceiling rose $17.2$\,pp and the image arm followed it up, landing at $0.3015$ --- essentially
the old ceiling. Purity rising alone would suggest collapse, but the categories in use
\emph{widened} at the same time ($102 \to 137$), which collapse does not do. So the blocker was
never ``we need Korean product images''; it was language, and an English catalog plus translation
raises the ceiling. Re-validating across $456$ verticals, the image-to-ceiling ratio actually
improves ($0.585 \to 0.687$) with the collapse diagnostic still negative; absolute values drop
because the ceiling itself is lower across a wider space.

\subsection{Wired In but Kept Disabled}

The offline-recipe axis is exhausted, and the remaining decisive question --- the difference
between having this on and off on the product surface --- belongs exclusively to the pilot. For
the pilot to measure it, the path must already exist in the SDK, so we wired it in. All of it,
however, ships disabled by default.

The projection artifact format carries a revision tag, and \code{project(x, device\_revision)}
checks it against the revision reported by the device OS and rejects a mismatch. The type system
blocks the path by which a $2048$-dimensional vector from iOS 16 could be silently run through
revision-2 weights. Quantization is effectively free: the minimum cosine similarity under
per-row int8 quantization is $0.999948$, with a mean of $0.999967$ over $256$ probes.

The projection output flows only into the content sketch, and reaching the ranking requires both
content consent and the user-embedding toggle to be on. The gate runs actual wasm to demonstrate
ten cases --- the consent ladder, absence of a projection, revision mismatch, revocation
cascades, and others.

\paragraph{The content channel is off by default.}
The consent tier is at its highest level (\S\ref{sec:arch}), the ranking-channel flag is false,
and the user embedding defaults to off. The default state requires all three independent
switches to be off simultaneously.

\paragraph{Layer 3 was blocked by the runtime.}
The OS language-model path was approved and actually exercised, but on the test device the
availability query returns ``unavailable'' rather than ``downloadable,'' and session creation
fails. That is, the download never even starts, and the blocker for Layer 3 is the runtime, not
approval. This was confirmed at zero cost.

\section{The Federated Personalization Path and Its Privacy Accounting}
\label{sec:fl}

\subsection{Round Protocol}

The server publishes a round identifier, the global parameters $\theta_t$, a deadline, and a
target number of participants $K$. Clients participate opportunistically (while charging, on
Wi-Fi, idle), download $\theta_t$, train for $E$ epochs on local data, form
$\Delta_i = \theta_i - \theta_t$, clip $\|\Delta_i\|_2$ to $C$, add noise, mask, and upload. The
server aggregates only what arrives by the deadline.

\begin{equation}
\theta_{t+1} = \theta_t + \frac{1}{n}\sum_{i \in \mathcal{A}_t} \Delta_i,
\qquad \mathcal{A}_t = \{i : \text{arrival time}_i \le \text{deadline}_t\}
\end{equation}

\begin{table}[H]
\centering\footnotesize
\caption{Federated protocol invariants.}
\begin{tabularx}{0.95\textwidth}{l L L}
\toprule
\textbf{\#} & \textbf{Invariant} & \textbf{Why} \\
\midrule
F1$^{*}$ & The server must not be able to see individual deltas & The privacy accounting in
\S\ref{sec:dp} applies only under this condition \\
F2 & Deltas are clipped before upload & Noise without clipping provides no guarantee \\
F3 & The deadline is itself the straggler policy & Waiting for slow devices would prevent the round from ever closing \\
F4 & Late-arriving deltas are discarded & Mixing in deltas computed against a stale $\theta$ causes silent divergence \\
F5 & A round is cancelled if participation falls below the minimum & Aggregating too few participants exposes individuals \\
F6 & Rounds are logged reproducibly & Without this, regressions cannot be caught \\
\bottomrule
\end{tabularx}
\end{table}

$^{*}$\,F1 is a \emph{requirement of the Stage-2 threat model}, not a property the current
system demonstrates. The masked aggregation prototype (\S\ref{sec:rw-fl}) cancels masks only
under full participation; without share-based recovery a dropped client leaves a mask standing.
Every $\varepsilon$ in \S\ref{sec:dp} is therefore conditional on F1, and we mark it rather than
let the accounting read as unconditional.

The dropout result in C5 must be read the same way. The 30\% dropout simulation runs the
aggregator's straggler policy \emph{with masking deliberately not applied}, because simulating
dropout under masking would be exercising a recovery mechanism this codebase does not implement.
Mask cancellation is proven separately, under full participation, by a different test. The two
properties --- ``the round survives stragglers'' and ``masks cancel'' --- are each demonstrated,
but not yet together.

Heterogeneity tolerance is not magic; it is three concrete implementations. Deadline-based
asynchronous collection avoids a synchronous barrier, per-device-tier participation caps prevent
bias toward high-end devices, and partial training is allowed but capped in its weighting by
completed epochs (because that weighting itself becomes a channel for high-end bias). Measured
results: at a $30\%$ dropout rate, round aggregation reaches $50/50$, with holdout BCE going from
$0.6936 \rightarrow 0.2125$.

\subsection{What It Means to Close the Loop}

The longest-surviving defect in this project had no symptoms. Uploads returned 202, the gate
passed, and the screen looked like rounds were running --- yet \emph{the device was receiving
nothing}. Three places were disconnected: the global parameters were initialized as a zero
vector, so what went out was not a delta but the absolute parameters trained from scratch each
round; the personal head was discarded immediately after being computed; and the aggregated
result arrived in the response, but nobody read it.

The lesson is that ``the round was accepted'' is not evidence that ``training actually ran.'' A
regression test now checks four things --- application of the aggregated result, movement of the
head, intervention by the primary ranker, and survival across a refresh --- and this test was
validated against a deliberately broken build.

The personal head intervenes only on the primary ranker's top 32 candidates. The reason is not
cost but explainability: if the personal head could promote a category the base ranker never
even considered, we could not answer ``why did this appear.'' It also adds a contribution
proportional to the spread of candidate scores, not an absolute value. Because the spread of
cosine scores differs by query, a fixed constant would do nothing for some queries and completely
upend the ranking for others.

\subsection{On-Device Labels}

The device has no ground-truth labels. Wiring up the combiner without deciding what counts as a
positive leaves a training path that exists but does not train anything.

\begin{table}[H]
\centering\footnotesize
\caption{Assignment of on-device training signals. Source \code{docs/spec/17-on-device-labels.md}.}
\begin{tabularx}{0.95\textwidth}{l L c c}
\toprule
\textbf{Signal} & \textbf{What it says} & \textbf{Combiner $\theta$} & \textbf{Personal head $w$} \\
\midrule
Creative tap & The user pointed to one leaf & Positive & Positive \\
Positive feedback & The whole answer was acceptable & Unused & Positive \\
Negative feedback & The whole answer was wrong & Unused & Negative \\
In-session requery & The prior answer was possibly wrong & Unused & Unused \\
\bottomrule
\end{tabularx}
\end{table}

The combiner's objective is a listwise softmax over the candidate set, so it requires knowing
``which one was it within this set,'' and the tap is the only signal that provides that
information. The personal head uses binary cross-entropy, so positives and negatives feed in
directly as given.

Three biases cannot be eliminated, so we state them explicitly. A tap is a choice among what we
showed, so a leaf that never made it into the results can never become a positive (implicit
feedback bias). We do not use our own classification output as a label --- feeding predictions
back in as labels amplifies the model's own confidence. And taps are rare.

Order also matters. A single creative tap both leaves a training sample and pushes an event into
the behavior window, and that tap event's category hint is exactly the ground-truth leaf. If the
event is pushed in first, the training sample learns by looking at its own answer. Reversing the
order still passes type checking, linting, and every other test, so a dedicated test pins the
order explicitly.

\subsection{Differential Privacy Accounting}
\label{sec:dp}

The $\varepsilon$ we had long been citing was \emph{per round}. The value that actually
corresponds to a person's privacy is the whole campaign. Composing 50 rounds gives $17.67$ by
hand-computed zCDP, which falls outside the Gboard production range
($0.99$\,\textasciitilde\,$13.7$) we cited as a reference on the very same screen. In other
words, our documentation was writing the favorable number large and deferring the unfavorable
one.

The cause was not the noise but a missing model term: the accounting had no \emph{device
population}. Computing subsampling amplification requires the per-round cohort divided by the
total population, but that parameter did not exist, so the calculation was implicitly using the
worst-case assumption that effectively every device participates in every round.

\begin{table}[H]
\centering\footnotesize
\caption{Privacy budget. Top: campaign-wide $\varepsilon$ (PLD accounting, 50 rounds,
$\delta = 10^{-10}$, effective noise multiplier $z = 3.162$). Bottom: per-round $\varepsilon$
versus number of participants. The minimum number of participants satisfying the target of
$5.0$ is $204$. Source \code{gates/dp\_gate.py}, \code{artifacts/ui-facts.json}.}
\begin{tabular}{l r}
\toprule
\textbf{Assumption} & \textbf{Campaign $\varepsilon$} \\
\midrule
No amplification ($q = 1$) & $16.25$ \ (hand-computed zCDP $17.67$) \\
Population $10{,}000$, cohort $1{,}000$ ($q = 0.1$) & $\mathbf{1.54}$ \\
Population $100{,}000$ ($q = 0.01$) & $0.14$ \\
Composed including the attribution channel & $\mathbf{1.7295} \le 2.0$ \\
\bottomrule
\end{tabular}

\vspace{4pt}
\begin{tabular}{l r r r r r r}
\toprule
Participants $n$ & $10$ & $50$ & $\mathbf{204}$ & $500$ & $1{,}000$ & $10{,}000$ \\
$\varepsilon$ (per round) & $26.46$ & $10.60$ & $\mathbf{4.996}$ & $3.13$ & $2.196$ & $0.684$ \\
\bottomrule
\end{tabular}
\end{table}

The accounting was not hand-rolled. We use Google's \code{dp-accounting}\cite{dpacct} PLD
implementation, cross-checking at $q = 1$: PLD gives $16.25$ against the hand-computed zCDP value
of $17.67$. PLD being tighter is correct, which confirms the wiring is right.

Amplification is not free. We carry its three preconditions all the way through the data to the
display screen: that participation is actually random, that Secure Aggregation holds (individual
updates are not observable) --- which, as \S\ref{sec:rw-fl} states, our prototype does not yet
provide --- and the Poisson-sampling assumption. Real federated learning is
sampling without replacement governed by device availability, which is not exactly the same, and
if the preconditions cannot be met, the no-amplification value must be used instead. The gate
judges against the campaign-wide value, and if the accounting library is unavailable it returns
``unmeasured'' rather than a pass.

\paragraph{$\varepsilon \le 2.0$ is not a technical parameter but a deployment precondition.}
Meeting the target $\varepsilon$ requires a device population of $10{,}000$ or more. Deployed on
an early-stage publisher with a few thousand installed devices, amplification does not hold, and
honesty requires using $16.25$ instead. We therefore split the rollout into two stages. In the
Stage 1 pilot, federated learning is turned off and we sell only on-device intent inference and
zero raw-text egress. With learning off, the privacy claim is not $\varepsilon$ but zero
raw-text egress, which is the stronger claim of the two. In Stage 2, once the population
precondition is met, federated learning is turned on and the campaign-wide $\varepsilon$ is
disclosed.

\subsection{Attribution Channel}

The campaign $\varepsilon$ of $1.54$ that we cited before 2026-08-26 was the value for the
federated-delta channel alone. The impression- and click-aggregation channels had no cap and no
accounting; the histogram was an unbounded list of raw ad identifiers. It is now bounded: $28$
buckets (equal to the number of top-level categories in the intent head), $3$ active buckets per
round, a per-bucket count cap of $1$, noise $\sigma = 0.3$, and a minimum bucket count of $50$.
The number that should actually be cited is $1.7295$, the composition of both channels; the
federated-delta channel alone accounts for $1.5422$ of it, so the attribution channel costs
$+0.187$. Until recently the gate did not produce that number at all: it returned early on the
axis mismatch below, so the composition was computed nowhere and the figure lived only in a
design document. The axis check and the composition are independent, so the gate now completes
the accounting and still fails on the axis.

The noise $\sigma$ was set by measured sweeps. $\sigma = 0.2$ gives a composed $\varepsilon$ of
$1.981$, so close to the target of $2.0$ that it would tempt the next person to just raise the
target to pass. $\sigma = 0.3$ gives $1.730$, leaving $13.5\%$ headroom, with a billing error of
about $\pm 67$ on roughly $2{,}500$ impressions per bucket at the cumulative campaign level, or
$2.7\%$. At the per-round level, however, the error is $19\%$, so per-round settlement is not
possible.

\subsection{The Personalization Axis Is the Weakest}

The user embedding is structured so that the ``large half'' is trained on a consented panel and
shipped, and only the small combiner is federated. The shipped weights total $52{,}736$
parameters ($W_h$ $284 \times 128$, $W_c$ $128 \times 128$), a file of $54{,}408$ bytes,
$210{,}944$ resident bytes, and a latency increase of $+0.342$\,ms. The device does not train
these weights --- \code{sdk-core} has no backpropagation.

The measured gain is not small.

\begin{table}[H]
\centering\footnotesize
\caption{User embedding arms. Panel $6{,}000$, holdout $5{,}997$, disjoint by person.
The baseline is the shipping build. The left two columns use uniformly interest-weighted
personas (\code{ship\_arch.json}); the right two columns use personas regenerated with the
2026-09-01 REES46 correction actually applied (\code{ship\_arch\_fix.json}, after fixing the
generator's binary-search defect).}
\begin{tabular}{l r r r r}
\toprule
& \multicolumn{2}{c}{Uniform weighting} & \multicolumn{2}{c}{\textbf{Correction applied}} \\
\textbf{Arm} & Mid-level (L2) top-5 & Sub-category top-1 & Mid-level (L2) top-5 & Sub-category top-1 \\
\midrule
B: Shipping build + personal head & $+6.19$\,pp $[5.55, 6.80]$ & $+3.74$\,pp & $+5.00$\,pp $[4.40, 5.62]$ & $\mathbf{-2.53}$\,pp \\
C: Personal head + federated & $+7.92$\,pp $[7.25, 8.64]$ & $+3.95$\,pp & $+6.94$\,pp $[6.27, 7.62]$ & $\mathbf{-1.75}$\,pp \\
D: Federated only & $+7.09$\,pp $[6.44, 7.75]$ & $+2.62$\,pp & $+3.57$\,pp $[3.04, 4.09]$ & $-5.14$\,pp \\
\bottomrule
\end{tabular}
\end{table}

With the correction applied, the conclusions for both halves hold and, if anything, strengthen
($C-B$: $+1.93$\,pp $[1.50, 2.37]$; $C-D$: $+3.37$\,pp $[2.85, 3.94]$). But \textbf{sub-category
top-1 falls below the shipping build in every arm}. For concentrated interests, the user arm
appears to lift the entire dominant block, gaining at the mid-level (L2) surface while
scrambling the leaf order within the block. Read under this paper's rule of judging both surfaces
simultaneously, this artifact fails the enable criterion on the corrected data.

The runtime default remains off nonetheless, and the verdict is explicitly ``do not enable.''
Neither budget nor code is what blocks it. L1 sits at $68.6\%$, L3 at $28.0\%$, latency adds
$+0.342$\,ms, and $\varepsilon$ is unchanged --- all with room to spare. What blocks it is the
fact that \textbf{there are only 3 real labels}.

\begin{table}[H]
\centering\footnotesize
\caption{Shortage of real labels. Source \code{experiments/user-vector-gap/default\_on\_verdict.json}.}
\begin{tabular}{l r r r}
\toprule
 & \textbf{Current} & \textbf{Needed} & \textbf{Shortfall factor} \\
\midrule
Creative-tap labels for training & $3$ & $800$ & roughly $250\times$ \\
Creative-tap labels for evaluation & $3$ & $1{,}750$ & roughly $580\times$ \\
\bottomrule
\end{tabular}
\end{table}

Enabling the default requires five preconditions: a real-panel training size of at least
$6{,}000$ people; a real-panel holdout of about $6{,}000$ people, disjoint by person; at least
$800$ real training labels and $1{,}750$ real evaluation labels; that \emph{both} mid-level (L2)
top-5 \emph{and} sub-category top-1 have a confidence-interval lower bound above 0; and
reproduction of a negative control (someone else's history).

The fourth condition matters because there is a measured precedent of sign reversal: within the
same trained-combiner family, mid-level (L2) top-5 rises significantly by $+0.80$\,pp
$[0.34, 1.31]$ while sub-category top-1 falls significantly by $-0.91$\,pp $[-1.71, -0.11]$.
Flipping a default on the strength of a single favorable granularity means nobody ever looks
again.

The same discipline applies to the weight constant. Sweeping the user-arm weight $\alpha$,
$\alpha = 2.0$ looks best on the surface at $+7.02$\,pp, but sub-category falls below baseline
at $-3.35$\,pp. The shipped value $\alpha = 1.0$ ($+6.02$ / $+3.75$) is the defensible one.

\section{The Ad Call --- OpenRTB 2.6 and Query-less Placements}
\label{sec:adcall}
\label{sec:rtb}

A ranking produced on the device only creates value for publishers and advertisers once it
becomes an ad call. We embedded this final leg using the standard protocol
(OpenRTB~2.6~\cite{openrtb}), but
enforced the privacy boundary through types rather than documentation.

\subsection{Closed Request Types}

The \code{user} object of \code{BidRequest} is defined as a struct with only two fields,
\code{data} and \code{ext}. Identifier fields such as \code{user.id} and \code{device.ifa} are
not \emph{omitted}; they are \emph{never declared} --- adding one would require changing the
type, and that change is caught by the sabotage check of the egress gate (the C3 extension).

The \code{ext} field deserves its own sentence, because in OpenRTB it is ordinarily an open
extension container and would defeat the whole argument: if \code{ext} were a free-form JSON
map, an identifier would simply live at \code{ext.foo.user\_id} and the closed \code{user}
struct would prove nothing. It is not one. \code{ext} has the concrete type \code{UserExt},
which contains exactly one field (the vendor namespace, written \code{oicr} throughout this
paper and in the published bundle), whose type in turn declares five fields and
nothing else. Every struct along that path carries \code{deny\_unknown\_fields}, so the
boundary holds in both directions: an unexpected key fails to serialize on the way out, and
fails to deserialize on the way in.
\code{device} is closed to 5 fields (devicetype, language, lmt, os, osv), and \code{lmt=1} is a
constant. Measured requests are $1{,}721$ bytes per call under the shipping-path classification,
with 5 intents, and all 7 raw-text canaries undetected. Intent is serialized in parallel through
two channels --- our extension (\code{user.ext.oicr} v1) and the standard segment
(\code{user.data}, segtax~7) --- so that even exchanges that do not read the extension still
receive the mid-level (L2) signal through the standard channel.

\subsection{Leaf Granularity and Sensitive Categories}

Intent-signal granularity comes in two layers. The standard segment carries the mid-level
category (L2, 252-bin), while only our extension carries the optional field \code{leaf\_id}
(6{,}020-bin). The granularity increase is about $+3.5$ bits per signal, but since requests
carry no identifier, cross-request linkage is bounded at the same axis as before --- the
server-side IP level. Sensitive categories \emph{drop} the leaf: top-level categories
\{healthcare, finance\} and mid-level categories \{pregnancy and childbirth, senior health\} are
sent only up to the L2 summary, and \code{None} means the key itself is absent from
serialization. On the matching side, a leaf creative is used with priority if present, but only
accepted when it falls within the same L2; otherwise matching falls back to the L2 pool.

\subsection{Korean Placement Specifications and the Creative Pipeline}

To keep protocol validation from remaining confined to screenless JSON, we embedded 11 display
placements from Naver GFA and Kakao (Bizboard, Moment) as a registry (\code{kr\_slots}),
preserving their exact pixel specifications and size caps. The request's
\code{imp.banner\{w,h\}} and \code{imp.tagid} designate the placement, and the exchange serves a
creative matching that pixel specification exactly. Creatives are, first and foremost, JPG/PNG
assets composited by cropping license-clear free stock photography (Openverse --- cc0, pdm, and
by licenses only) to the placement dimensions, with every asset's source URL, author, and license
recorded in a ledger (\code{LICENSES.json}) --- an image with no record is not used. The size cap
is enforced by an \code{assert} at the compositing stage. A creative with no real photograph
falls back to SVG, but the response extension honestly marks it as
\code{asset = photo | synthetic-svg} --- there is no path by which the fallback masquerades as a
real photograph (a request for an unregistered raster returns 404).

\subsection{Query-less Placements --- Ads Served Without an Utterance}

Most display ads are served at moments when the user is not searching. When there is no
utterance, \code{classify\_context} produces the ranking: it combines the direction of the
folded content sketch (articles the user has read --- perception layer 1 of \S\ref{sec:percept}),
the profile prior, and the three behavior-window signals to rank leaves, and that ranking then
enters the \emph{same} request builder as the utterance path. The egress contract (the closed
user type, the confidence floor) is identical. The consent ladder is a precondition --- without
content consent there is no sketch, and if the sketch is empty the system raises an error rather
than fabricating a ranking. The full text of articles read never appears in the request; a web
E2E test checks for the article body's distinctive phrases as canaries.

History alone can also stand a request. Because the additive prior is by design unable to dominate the text term (\S\ref{sec:match}), when the folded content carries no commercial signal and confidence falls below the floor, a \emph{history-segment fallback} fires: a ranking built only from the profile's top-level history and behavioral hints (mid-level and leaf granularity) enters the same request builder. This is precisely the shape of display retargeting: reading the same essay, a user with a running history receives the running-shoe creative and a user with a childcare history receives the car-seat creative (the E2E suite pins this swap). With no history, no segment is fabricated --- no ad is served.

Images are also a signal for these placements. The OS embedding (rev2, 768-dimensional) of an
image the user attended to is folded into the content sketch through a rectangular projection
$W \in \mathbb{R}^{768 \times 128}$ --- neither the original image nor the raw embedding is
stored; both are projected and discarded immediately. This projection is a contrastive-learning
candidate from \S\ref{sec:percept} and is \textbf{not yet a shipping artifact}. The demo loads
the candidate through the same entry point designed for a pilot A/B to carry a candidate $W$, and
the screen states, in plain language, ``research candidate --- pre-deployment gate.'' Because
browsers have no standard OS vision API, image embeddings for the web demo were precomputed with
macOS Vision, while the iOS verification app calls the same API live.

\subsection{Validation and Limitations}

On the server side, a mock exchange (accept 200 / no-bid 204 / reject 400 / overload 503, with
win- and billing-notice counts) closes the round trip, and a web E2E test using the real wasm and
a real server checks slot serving, absence of raw text, the notification path,
placement-specification switching, leaf-priority matching, and query-less placements (including
rejection when consent is absent). Limitations to state explicitly: we have not connected to a
real exchange (a URL swap and \code{test=0} remain to be done); the mock exchange's matching
quality says nothing about a real exchange's; and \code{nurl}/\code{burl} structurally let the
exchange learn that it won, while IP is an axis the SDK cannot erase.

\section{Purchase Probability as a Ranking Signal}
\label{sec:purchase}

Intent tells us \emph{what} a person is looking at. It does not tell us whether they are about
to buy it, and those are different questions: someone who bought coffee beans yesterday is
reading about coffee beans for a different reason than someone who has not bought any in three
months. This section reports the signal we add for that, what it is worth on a real purchase
log, and what is still missing. It also reports a measurement mistake of our own, because the
mistake is the more transferable result.

\subsection{A Closed-Form Hazard, Not a Learned Model}

The device cannot host a gradient-boosted ensemble per category. The budget is roughly ten bytes
per leaf, an $O(1)$ update per observed event, and no inference runtime at all --- the same
constraint that forced a static embedding table in \S\ref{sec:distill}. We therefore model
repurchase as a hazard over time since the last purchase in that category, with a per-category
cycle prior and a count term, evaluated as arithmetic rather than inference. We call it
\code{mPG}. The form is not ours: buy-till-you-die models in marketing science have described
repeat purchase this way for two decades, and the BG/NBD formulation is popular precisely
because its parameters fall out in closed form rather than needing an optimizer~\cite{bgnbd}.
That property is usually a convenience; for us it is the requirement. Whether that trade costs accuracy is an empirical question, and \S\ref{sec:pb-ceiling}
answers it.

\subsection{A Degenerate Evaluation Task Looked Like a Weak Model}

Our first measurement on a real e-commerce behaviour log (REES46 mirror, 34.8\,M rows over 21
days) returned AUCs of $0.53$--$0.59$, against $0.85$--$0.93$ on synthetic sequences drawn from
our own hazard assumptions. The obvious reading is that the synthetic numbers were optimistic
and the model is weak on real behaviour. That reading was half wrong, and diagnosing the
distribution before touching the model is what showed it.

\begin{table}[H]
\centering\footnotesize
\caption{Diagnosis of the first harness. Each row is a property of the evaluation task, not of
the model. Source \code{experiments/purchase-lift/diagnose\_real.py}.}
\begin{tabular}{l l}
\toprule
\textbf{Observation} & \textbf{Consequence} \\
\midrule
$62.6\%$ of evaluation rows had a single prior purchase & $p_{\text{buy}}$ returns the same
constant for all of them \\
Positive rate $90.4\%$ ($w=7$d), $46.5\%$ forced to 1 & the window truncation is being measured,
not the model \\
Median inter-purchase gap $0.5$ days & one order split across items counted as a repurchase \\
$79.8\%$ of pairs dropped for having no repurchase & censored users removed from the evaluation
\\
\bottomrule
\end{tabular}
\end{table}

Each of these caps the achievable AUC independently of the predictor. The fourth is the one a
survival formulation exists to avoid: a customer who has not repurchased yet is censored, not
absent, and web-scale repurchase work models exactly that rather than discarding
it~\cite{walmart2026}. The second harness (\code{eval\_real\_v2.py}) folds
events to orders (\code{user\_session}), fixes a snapshot $t_0 = t_{\max} - w$, admits
\emph{every} $(\text{user}, \text{category})$ with at least one purchase before $t_0$, and labels
by purchase in $(t_0, t_0 + w]$ --- a horizon entirely inside the observation window, so nothing
is censored and recency becomes a live axis rather than a constant.

\textbf{The same deployed formula, with no change to the model, scores $0.7425$ ($w=3$d) and
$0.6915$ ($w=7$d).} We do not claim this as an improvement over $0.53$--$0.59$: the two harnesses
ask different questions (``among people who repurchased, who comes back sooner'' versus ``will
someone who has bought before buy again within $w$ days''), and comparing them directly would
repeat the error this subsection is about. What we claim is narrower and more useful --- half of
what looked like model weakness was the evaluation, and we would not have found that by improving
the model.

\subsection{Where the Signal Comes From, and Its Ceiling}
\label{sec:pb-ceiling}

An ablation over the second harness (user-disjoint split, bootstrap 2{,}000) separates the
contributions.

\begin{table}[H]
\centering\footnotesize
\caption{Ablation on real logs. $\Delta$ is against the deployed formula; brackets are 95\%
bootstrap CIs. Source \code{experiments/purchase-lift/real-v2-ablation-2026-09-05.json}.}
\begin{tabular}{l r l r l}
\toprule
\textbf{Arm} & \textbf{$w{=}3$d} & \textbf{$\Delta$} & \textbf{$w{=}7$d} & \textbf{$\Delta$} \\
\midrule
\code{mPG} (deployed)            & $0.7327$ & --- & $0.7000$ & --- \\
Weibull $m{=}0.9$                & $0.7385$ & $+0.0058$\,[$+0.0032,+0.0082$] & $0.7026$ & $+0.0026$\,[$+0.0012,+0.0038$] \\
\midrule
\ding{172} L2 base rate only     & $0.5755$ & $-0.1572$ & $0.6033$ & $-0.0967$ \\
\ding{173} $+$ frequency $k$     & $0.6885$ & $-0.0442$ & $0.6826$ & $-0.0174$ \\
\ding{174} $+$ purchase recency  & $0.7454$ & $+0.0127$\,[$+0.0057,+0.0196$] & $0.7249$ & $+0.0249$\,[$+0.0162,+0.0333$] \\
\ding{175} $+$ Weibull logit     & $0.7456$ & $+0.0129$ & $0.7257$ & $+0.0257$ \\
\ding{176} $+$ view / cart       & $0.7563$ & $+0.0236$\,[$+0.0152,+0.0328$] & $0.7336$ & $+0.0336$\,[$+0.0231,+0.0434$] \\
\bottomrule
\end{tabular}
\end{table}

Almost all of the gain is \ding{174}: letting base rate, frequency and recency enter with
\emph{free weights}. Adding the closed-form Weibull logit on top of them (\ding{175}) contributes
essentially nothing ($+0.0002 / +0.0008$), and its fitted coefficient is \emph{negative}
($-0.65 / -1.49$). The interpretation is specific: the negative-binomial form~\cite{bgnbd} fixes
those three quantities into the combination $(\beta+T)/(\alpha+k)$, and once the three are available
separately, that fixed combination is a liability the fusion subtracts back out.

One more result falls out of the same sweep: the best-fitting shape parameter is $m < 1$
($0.85$--$0.9$), i.e.\ the marginal hazard \emph{decreases} with time since purchase. That
matches the independently reported $\hat{k} = 0.911$ in web-scale repurchase
work~\cite{walmart2026}, which is reassuring --- and it is the opposite of what our
recently-purchased suppression rule assumes. We keep the suppression, because it is a business
rule (``do not re-advertise what was just bought'') rather than a prediction, but the two must
not be described as the same model.

The ceiling is set by information, not by model capacity. \textbf{LightGBM given the same
features reaches $0.7316 / 0.6996$, against the deployed closed form's $0.7425 / 0.6915$} --- a
tie. The arithmetic hazard is already at the ceiling of what these features support, so the
deployability trade in \S3 costs nothing measurable here. What lifts the ceiling is more
information (\ding{176}, browse and cart signals) or free weights (\ding{174}), not a bigger model.

\subsection{Does It Survive Time, Our Own Taxonomy, and the Deployed Code?}

Three checks, because a coefficient fitted and evaluated on one snapshot is not evidence that a
federated round learned in the past will help in the future.

\textbf{Time.} Fitting coefficients on an early snapshot's population and evaluating on a
snapshot seven days later with zero user overlap, \ding{174} retains $88$--$89\%$ of its gain
($+0.0112$\,[$+0.0076,+0.0149$] at $w{=}3$d; $+0.0222$\,[$+0.0180,+0.0262$] at $w{=}7$d), and the
fitted coefficients are stable across four fits (\code{log\_cat\_rate} $0.92$--$1.09$ against a
theoretical $1.0$). \ding{176} does not survive: its marginal contribution over \ding{174}
collapses from $+0.0108/+0.0087$ to $+0.0009/+0.0032$. Browse and cart coefficients do not cross
time, which is why we did \emph{not} add per-leaf view and cart timestamps to device state --- a
measurement that demoted our own proposal.

\textbf{Our taxonomy.} REES46's 132 categories are not our category system. Bridging them onto
our mid-level categories and re-running the same ablation over 71 of ours gives $\Delta$
$+0.0040$\,[$+0.0014,+0.0064$] ($w{=}3$d) and $+0.0016$\,[$+0.0008,+0.0023$] ($w{=}7$d) --- the
same size, CIs excluding zero. Absolute AUC drops slightly ($-0.010/-0.004$) because 132 bins
became 71, not because the method weakened.

\textbf{The deployed code.} The Python harness measured the lift; the Rust path is what ships.
Replaying the same evaluation rows through \code{PurchaseView::rank\_score} reproduces the Python
figure exactly: $0.7359$ against $0.7359$ ($w{=}3$d) and $0.6900$ against $0.6900$ ($w{=}7$d).

\subsection{Combining It With Intent}

The ranking score adds the purchase term with weight $\lambda_p$. We swept it on real histories
through the deployed boost function itself, over $26{,}194$ $(\text{user}, \text{category})$
pairs.

The sweep contains a trap worth reporting. Ranking a category only against the same user's
history \emph{in that category} --- the optimistic setting --- the gain rises monotonically to
$\lambda_p = 2.0$ ($+2.0$\,pp top-1), and never bends, so it cannot select a value at all. Once
other categories compete for the same slot, which is what deployment does, the ordering reverses:
$\lambda_p = 2.0$ is \emph{significantly worse} than the deployed default ($-1.55$\,pp, 95\% CI
$[-1.83,-1.26]$) while $\lambda_p = 1.0$ stays significantly better ($+1.73$\,pp,
$[+1.57,+1.90]$). A fine grid puts the optimum at $1.0$ with a $0.9$--$1.1$ plateau. Had we
selected on the optimistic setting we would have shipped the one value the honest setting
rejects.

\subsection{Learning the Combiner Federated --- and What Is Missing}

The combiner coefficients are learned in the same federated rounds as everything else, from
labels the device derives locally: a decision is observed, confirmed positive if a purchase
follows, expired negative otherwise. In a 50-round simulated campaign the rank head converges
from AUC $0.5$ to $0.7734$, and the four added coordinates leave the whole-campaign privacy
budget unchanged at $\varepsilon \approx 1.54$, inside the $\varepsilon \le 2.0$ cap.

Turning the loop surfaced two guards. A single browser cannot form a round --- the
$k$-anonymity floor is 8 participants --- so the demonstration path fills the cohort with real
submissions rather than synthetic peers, whose contribution is exactly zero. And because the
aggregate carries DP noise, coefficients become non-zero after the first round whether or not
anyone trained; a threshold at $0.106 = 3\sigma/\sqrt{N_{\min}}$ separates the two, set after
confirming that trained and noise-only magnitudes actually separate ($0.035$ versus
$1.29$--$2.29$).

\textbf{What is missing is the coefficients themselves.} The path exists end to end and is
exercised by convergence tests, Rust/Python parity, and browser E2E --- but coefficients come
from a pilot cohort of real people, which we do not have. Without them the ranker falls back to
the hazard path silently and correctly, forgoing the measured gain. The cold-start prior is
likewise unbuilt: a 21-day window collapses the per-category cycle distribution (median gap
across mid-level categories is 2 days, carrying no information between categories), and we keep
the diagnostic with the script rather than rebuild it until a log spanning months exists. We
state both gaps instead of filling them with synthetic coefficients, which would be
indistinguishable in the paper and wrong in deployment.

\section{Experimental Results}
\label{sec:results}

\subsection{Real-Data Korean Benchmarks}

The benchmarks in this subsection sit beside, not instead of, the external Korean commerce
evaluation of \S\ref{sec:aihub}; they measure the same encoder on a task other people defined. So we
separately measured the same axis --- classification over short Korean text --- on real data and
real labels made by others: KLUE-YNAT\cite{klue} (7-way news-headline classification) and
NSMC\cite{nsmc} (binary sentiment on colloquial movie reviews), with $40{,}000$ training /
$9{,}000$ evaluation examples, $2{,}000$ bootstrap resamples, and both training and evaluation
done with the embeddings from the \emph{shipping Rust core}.

\begin{table}[H]
\centering\footnotesize
\caption{Real-data Korean classification. Brackets show the 95\% bootstrap confidence interval.
Source \code{experiments/real-ko-bench/results.json}.}
\begin{tabular}{l r r}
\toprule
\textbf{Arm} & \textbf{KLUE-YNAT (7-way)} & \textbf{NSMC (binary sentiment)} \\
\midrule
Stage 1 (distill only) & $0.5240$ $[0.514, 0.535]$ & $0.7283$ $[0.720, 0.737]$ \\
\textbf{Stage 2 (+ tokenlearn)} & $\mathbf{0.6408}$ $[0.631, 0.651]$
  & $\mathbf{0.7362}$ $[0.727, 0.745]$ \\
Public Korean static embedding (64d) & $0.3987$ $[0.389, 0.409]$ & $0.6844$ $[0.676, 0.694]$ \\
Teacher, 768d (upper bound) & $0.8188$ & $0.8498$ \\
\midrule
\textbf{Shipping build} (128d, 4-bit; \S\ref{sec:gates}) & $\mathbf{0.6750}$ & $\mathbf{0.7548}$ \\
Retention relative to teacher (Stage 2) & $78.3\%$ & $86.6\%$ \\
Retention relative to teacher (shipping build) & $82.4\%$ & $88.8\%$ \\
\bottomrule
\end{tabular}
\end{table}

Three things stand out.

\textbf{First, candidate selection reproduced on real data.} The public static embedding beat us
by $5.2$\,pp on KLUE-STS, and so we had flagged, at medium confidence, that ``the proxy metric
and the product task diverge.'' On real-data classification, though, the public model loses
badly --- $-24.2$\,pp on YNAT and $-5.2$\,pp on NSMC (confidence intervals non-overlapping on
both). What diverged was not synthetic versus real data, but \emph{semantic similarity versus
classification}. The public model was built to directly target similarity from
sentence-entailment pairs; what we sell is classification.

\textbf{Second, tokenlearn was confirmed only by half.} It wins decisively on YNAT at
$+11.7$\,pp, but on NSMC it is $+0.8$\,pp with overlapping confidence intervals. That is, what
tokenlearn earns is topic and lexical discrimination, not sentiment polarity. We keep it, but we
do not write that it improves every task.

\textbf{Third, we can now state what compressing the encoder to $2.14$\,MB cost, on real data.} The shipping
build retains $82.4\%$ and $88.8\%$ of the teacher's performance (YNAT, NSMC) at $196.6\times$
smaller size --- the stage-2 row above is a training-stage ablation and retains less ($78.3\%$,
$86.6\%$), which is why the two pairs of figures differ. This compression ratio
compares the raw bytes of the embedding artifact ($2{,}141{,}820$) against the teacher checkpoint
($421$\,MB) on the same axis. We do not use the teacher directory's full size, since that would
include multiple copies of the same weights and inflate the ratio.

KLUE-STS scores $0.6031$ on the shipping path, versus $0.6112$ for a Python implementation of the
same weights. The gap of $-0.008$ is the cost of quantizing the embedding table, within the $0.02$
threshold.

\subsection{Real Korean Commerce Text Against Our Own Taxonomy}
\label{sec:aihub}

The benchmarks above are real, but they are not \emph{our} task: they classify news headlines and
movie reviews, not commerce text into our 256 mid-categories. That gap --- every accuracy figure
over our own taxonomy being generated by us --- was the single largest hole in this paper. We
closed part of it by applying for, and receiving, access to three Korean commerce corpora
published by AI-Hub (Korea NIA)~\cite{aihubabsa,aihubsmb,aihubcall}.\footnote{AI-Hub datasets 71603, 102 and 98, constructed under projects of
the National Information Society Agency (NIA) of Korea. AI-Hub's operators state that models and weights
produced by training on the data may be published without restriction, but that data merely
extracted, edited or otherwise re-processed from the originals may not be published, shared or
redistributed even with attribution, and that storing such material on overseas platforms would
itself constitute export. We therefore publish no corpus rows, no extracted term lists and no
quoted utterances anywhere in this paper or its artifact bundle --- only the category mapping we
wrote, the harness code, and aggregate measurements. Every figure below is reproducible by a reader
who obtains their own access and runs \code{experiments/real-ko-commerce/}.}

\paragraph{What the corpora are.} Dataset 71603 (\emph{aspect-based sentiment}) provides
$22{,}900$ labelled shopping-mall reviews across $5$ domains and $20$ categories, each carrying
the real product name it was written about. Dataset 102 (\emph{small-business customer order
Q\&A}) provides $307{,}551$ real customer utterances across $14$ categories, with
annotator-marked product-name entities --- these are order and enquiry utterances, the closest
public match to the input our system actually receives. Dataset 98 (call-centre Q\&A) turned out
\emph{not} to support an accuracy measurement at all: it carries no per-utterance product
category, so we use it only for lexical coverage and say so.

\paragraph{How the labels were mapped, and the control that makes it honest.} Neither corpus uses
our taxonomy, so we hand-built a mapping from their categories onto sets of ours, deciding each
entry by reading that category's actual product names and recording the rationale in
\code{aihub\_map.json}. A mapping like this is a judgement, and a broad one makes a top-5 hit
artificially easy: AI-Hub's ``daily goods'' bucket alone spans diffusers, pet food, tents and
strollers. We therefore report, beside every figure, a \textbf{permutation baseline} --- the same
mapping, the same label distribution, with only the link between query and label severed. The
gap between the two is what the model actually contributed. We also drop, rather than force, the
categories our taxonomy has no place for (AI-Hub's ``other'' bucket of flowers, bread and
scissors).

\begin{table}[H]
\centering\footnotesize
\caption{Deployment path on real Korean commerce text. Brackets are 95\% bootstrap confidence
intervals over $2{,}000$ resamples; \emph{perm.} is the permutation baseline under the identical
mapping, averaged over $400$ draws. Sources \code{experiments/real-ko-commerce/aihub\_surface.json} and
\code{aihub102\_surface.json}.}
\label{tab:aihub}
\begin{tabular}{l r r r}
\toprule
\textbf{Query set} & \textbf{$n$} & \textbf{Accuracy} & \textbf{perm.} \\
\midrule
\multicolumn{4}{l}{\emph{Dataset 71603 --- mid-category (L2) top-5, the product surface}} \\
Real product names, all & $1{,}174$ & $0.7504$ $[0.726, 0.775]$ & $0.1843$ \\
\quad --- leaf-name anchor present & $915$ & $\mathbf{0.8350}$ & --- \\
\quad --- anchor absent & $259$ & $0.4517$ & --- \\
\quad --- anchor absent, model-code form & $137$ & $0.2847$ & $0.1554$ \\
Real review sentences & $6{,}000$ & $0.4552$ $[0.4423, 0.4677]$ & $0.1241$ \\
\midrule
\multicolumn{4}{l}{\emph{Dataset 102 --- top-category (L1) top-5; its labels do not support L2}} \\
Product-name entities & $8{,}000$ & $0.7229$ $[0.713, 0.733]$ & $0.4785$ \\
Real customer utterances & $8{,}000$ & $0.5897$ $[0.579, 0.601]$ & $0.3813$ \\
\bottomrule
\end{tabular}
\end{table}

\paragraph{What each dataset can and cannot settle.} This paper now leans on five bodies of
evidence with different weaknesses, and it is easy to lose track of which one licenses which
claim. Table~\ref{tab:ladder} states that explicitly.

\begin{table}[H]
\centering\footnotesize
\caption{Evidence scope. No row is strong on every axis; the argument rests on their weaknesses
being different from one another.}
\label{tab:ladder}
\begin{tabularx}{\textwidth}{l L l}
\toprule
\textbf{Evidence} & \textbf{What it can settle} & \textbf{What it cannot} \\
\midrule
Synthetic held-out utterances & Task-aligned accuracy over our taxonomy, regression detection
  & Authored by us; not a real query distribution \\
ESCI queries translated to Korean & Behaviour on real production search queries
  & Amazon's user distribution in translation; labels derived, not human \\
AI-Hub 71603 & Real Korean product and review text with externally assigned source labels
  & Not queries; crosswalk to our taxonomy is ours \\
AI-Hub 102 & Real Korean customer utterances, $307{,}551$ of them
  & Labels support only the top category, not the product surface \\
Naver Shopping reviews & The distribution of real Korean commerce text
  & No labels at all; supports distribution comparison only \\
\bottomrule
\end{tabularx}
\end{table}

\paragraph{Validating the harness before reading it.} A measurement that agrees with our
expectations is the easiest kind to get wrong, and this repository has twice shipped a harness
whose numbers looked reasonable and were produced by a bug. So before interpreting anything we
ran three checks, recorded in \code{verify\_harness.json}. \emph{Truncation}: rolling $32$ leaves
up to mid-categories yields a median of $15$ distinct ones, and fewer than five for $0.43\%$ of
queries, so ``top-5'' is not silently a shorter list. \emph{Baseline stability}: a single
permutation draw is noisy --- our first one returned $0.2010$, which turned out to sit at the
upper edge of the $400$-draw interval $[0.1627, 0.2027]$; every baseline below is now a mean over
$400$ draws. \emph{Negative control}: scrambling the characters of each query drops mid-category
top-5 from $0.7504$ to $0.4736$, far outside the baseline interval, confirming the path reads the
query. That the scrambled figure still clears the baseline is itself informative: with mean
pooling, the unordered presence of sub-word units alone carries much of the signal.

\textbf{First, the product surface survives contact with real product names.} On the $1{,}174$
real product names, mid-category top-5 is $0.7504$ $[0.726, 0.775]$ against a mapping-preserving
null of $0.1843$, rising to $0.8350$ on the subset containing some leaf name as a substring.
That establishes that the deployed path transfers non-trivially to independently authored Korean
commerce text. It does \emph{not} establish our synthetic $77.6\%$ as an unbiased estimate of
accuracy on real search queries: the two sets differ in construction, in distribution and in what
a ``query'' is, and a figure landing between two others is not evidence that either is calibrated.

\textbf{Second, one observable separates success from failure, and it is not the one we first
reported.} Our first pass split these queries by whether the name contains a manufacturer model
code. A cross-tabulation then suggested two factors moving independently, and we reported them
that way. Controlling for which source category a query came from dissolves one of the two. What
survives is the \textbf{leaf-name lexical anchor}: whether the query string contains some leaf
name as a substring. We state the test operationally because the name invites a stronger reading
than the implementation supports --- it is not a morphological analysis. We strip whitespace from
the query and from every leaf name of two characters or more, and ask whether any leaf name occurs
as a substring of the query. That is precisely the condition under which the lexical-overlap term
of \S\ref{sec:matching} can contribute a non-zero score.

\begin{table}[H]
\centering\footnotesize
\caption{Mid-category top-5 on real product names, cross-tabulated, each cell against \emph{its
own} matched null. The null permutes labels \emph{within the cell}, so it preserves that cell's
label composition; $p$ is the fraction of $400$ permutations reaching the observed value. Source
\code{experiments/real-ko-commerce/p1\_stats.json}.}
\label{tab:aihub2x2}
\begin{tabular}{l r r r}
\toprule
\textbf{Cell} & \textbf{$n$} & \textbf{top-5} & \textbf{cell-matched null} \\
\midrule
Anchor present, no model code & $684$ & $0.8772$ & $0.2793$ \;($+59.8$\,pp, $p<0.0025$) \\
Anchor absent, no model code & $122$ & $0.6393$ & $0.1610$ \;($+47.8$\,pp, $p<0.0025$) \\
Anchor present, model code & $231$ & $0.7100$ & $0.2363$ \;($+47.4$\,pp, $p<0.0025$) \\
Anchor absent, model code & $137$ & $\mathbf{0.2847}$ & $0.1554$ \;($+12.9$\,pp, $p<0.0025$) \\
\midrule
Marginal, anchor present & $915$ & $0.8350$ & --- \\
Marginal, anchor absent & $259$ & $0.4517$ & --- \\
\bottomrule
\end{tabular}
\end{table}

The per-cell nulls correct a claim we made in an earlier draft. We wrote that the worst cell
``degrades toward chance.'' It does not. Chance \emph{for that cell}, under its own label
composition, is $0.1554$, and $0.2847$ sits $12.9$\,pp above it with no permutation out of $400$
reaching the observed value. The honest statement is that the worst cell loses most of the margin
the other three enjoy while remaining above its own chance level --- which is a boundary of
usefulness, not a collapse into noise.

\paragraph{The second axis does not survive a domain control.} Table~\ref{tab:aihub2x2} is a
cross-tabulation, and a cross-tabulation cannot tell a factor from its context. AI-Hub's twenty
source categories differ enormously in difficulty --- from $0.035$ to $1.000$ --- and they differ
just as much in how their products are named. So we regressed correctness on the two indicators
and their interaction, with and without source-category fixed effects. Because several categories
are perfectly separated, a logistic fit is singular here; we use a linear probability model with
HC1 robust standard errors, whose coefficients read directly in percentage points.

\begin{table}[H]
\centering\footnotesize
\caption{Correctness regressed on the two indicators, $n=1{,}174$, linear probability model with
HC1 standard errors. Adding source-category fixed effects leaves the anchor effect standing and
removes the model-code effect. The last column is an assumption-free check: the
category-size-weighted within-category difference, over the categories that contain at least five
queries on each side. Source \code{p1\_stats.json}.}
\begin{tabular}{l r r r}
\toprule
\textbf{Term} & \textbf{No fixed effects} & \textbf{$+$ source-category FE} & \textbf{Stratified} \\
\midrule
Lexical anchor & $+23.78$\,pp ($p{<}10^{-6}$) & $\mathbf{+13.01}$\,\textbf{pp} ($p=0.003$)
  & $+15.73$\,pp \;(16/20 cat.) \\
Model code & $-35.47$\,pp ($p{<}10^{-8}$) & $-4.01$\,pp ($p=0.51$)
  & $+0.20$\,pp \;(10/20 cat.) \\
Anchor $\times$ model code & $+18.74$\,pp ($p=0.005$) & $+4.59$\,pp ($p=0.48$) & --- \\
\midrule
$R^2$ & $0.194$ & $0.424$ & --- \\
\bottomrule
\end{tabular}
\end{table}

The model-code coefficient does not merely lose significance; its point estimate moves from
$-35.47$ to $-4.01$\,pp, and the model-free stratified difference is $+0.20$\,pp. That is
attenuation, not a standard error inflated by collinearity. The confound is visible in the data:
in the four fashion categories not one product name in $391$ carries a model code, while in the
four appliance categories between $46\%$ and $83\%$ do. The indicator we thought was measuring
surface form was in large part measuring \emph{fashion versus appliances}. We therefore withdraw
the two-axis claim. One axis --- the lexical anchor --- survives the control at roughly half its
uncontrolled size and remains significant. The model-code indicator adds nothing once the source
category is known, and we keep the cross-tabulation only as a description of where the failures
sit, not as a factorization of their causes.

What the failures do have in common is visible without a regression, and it is a property of the
input rather than of surface form: the names that fail consist of a brand sub-name, the maker's
name and an alphanumeric model code, with \emph{no category noun anywhere in the string}. A
static table over a $31{,}497$-token Korean vocabulary has no representation for such a string,
and the lexical channel has nothing to anchor on either. The boundary of applicability we should
have stated is therefore about information, not formatting: \emph{a listing whose title names the
manufacturer and the model but never the kind of thing it is carries no signal this system can
read}. Two of the twenty source categories fail this way badly enough to fall to their own chance
level or below --- one scores $0.035$ against a matched null of $0.162$, the other $0.036$ against
$0.040$ --- and a below-chance score is not noise but systematic misrouting, in this case toward
digital-device mid-categories whose leaf names are themselves alphanumeric.

\emph{A confound we nearly reported as a cause, and how we caught the second one.} AI-Hub masks
brand names in the fashion, cosmetics and household domains but not in appliances and IT. Split
on masking alone, the masked queries score \emph{higher} ($0.8803$ versus $0.4068$), which invites
the conclusion that removing brand tokens helps. It does not: masking is a domain indicator in
disguise, and it separates fashion and cosmetics from appliances and IT. We labelled that figure
a confound in the artifact and drew no causal reading from it --- and then reported the model-code
split as a mechanism, when it was the same confound wearing different clothes. The lesson we would
pass on is procedural: in a corpus whose categories were collected separately, \emph{any} query
attribute that varies by category is a domain indicator until a within-category comparison says
otherwise, and the cheapest defence is to run that comparison for every attribute you intend to
call a mechanism, not only for the one that looks suspicious.

\textbf{Third, conversational text is a different task and scores like one.} Review sentences
reach $0.4552$ $[0.4423, 0.4677]$ against a baseline of $0.1241$, and full customer utterances
$0.5897$ against $0.3813$. The same anchor axis governs them --- $0.5185$ versus $0.3492$
on reviews --- but a review carries sentiment about a purchase already made, and an enquiry
carries a transactional intent such as delivery timing or exchange, with the product often named
only obliquely or not at all. We report them as out-of-distribution stress tests --- neither an
upper nor a lower bound on the intended short-query task, but a measurement of what happens when
the utterance surface is wider than the one we designed for. Their granularities differ too:
$0.4552$ is a mid-category figure and $0.5897$ a top-category one, and the latter sits against a
null of $0.3813$, so it should never be quoted on its own.

\paragraph{Does the crosswalk make the result?} A hand-built mapping is ordinary evaluation
methodology, but ours has a defect we cannot undo: it was written while reading the same product
names we then evaluated on. That is not training leakage, it is \emph{evaluation-definition}
leakage, and the permutation null does not test for it --- a null preserves the mapping, so it
cannot ask whether an ambiguous mapping decision was resolved in our favour. Since the mapping
cannot be retroactively un-seen, we bound its influence instead: we narrow it and see whether the
conclusion survives. The narrowing uses the corpus's \emph{training} split, which the evaluation
never touches --- an AI-Hub category's mapped mid-category is kept only if some training product
name in that category contains a leaf name belonging to it.

\begin{table}[H]
\centering\footnotesize
\caption{Sensitivity of the result to how generous the crosswalk is. Lift is accuracy minus the
mapping-preserving null, itself recomputed for each mapping. Source
\code{experiments/real-ko-commerce/map\_sensitivity.json}.}
\label{tab:mapsens}
\begin{tabular}{l r r r r}
\toprule
\textbf{Crosswalk} & \textbf{L2 per label} & \textbf{top-5} & \textbf{null} & \textbf{lift} \\
\midrule
Hand-built (as reported above) & $2.55$ & $0.7504$ & $0.1843$ & $+56.6$\,pp \\
Kept only where the training split supports it & $1.60$ & $0.6882$ & $0.1576$ & $+53.1$\,pp \\
Single best-supported mid-category per label & $1.00$ & $0.5571$ & $0.1360$ & $+42.1$\,pp \\
\midrule
\multicolumn{5}{l}{\emph{Same narrowed crosswalk, restricted to products absent from the training split}} \\
Products unseen in training ($n=837$) & $1.60$ & $0.7838$ & $0.2376$ & $+54.6$\,pp \\
\bottomrule
\end{tabular}
\end{table}

Collapsing the crosswalk from $2.55$ mid-categories per label to exactly one moves absolute
accuracy a great deal --- $0.7504$ to $0.5571$ --- and moves the lift over the matching null much
less, from $+56.6$ to $+42.1$ points. We are careful about what that licenses. Every arm in this
table is a descendant of the same originally test-informed crosswalk: narrowing it removes
generosity but cannot remove the fact that its rows were written while reading the evaluation
product names. The defensible reading is therefore \emph{the qualitative transfer conclusion
survives progressively narrower descendants of the same crosswalk}, not that leakage has been
eliminated. Eliminating it would take a crosswalk built by someone who never saw the evaluation
names, which \S\ref{sec:limits} lists as outstanding; it is only twenty mappings, so the obstacle
is a second annotator rather than effort. Any single absolute number quoted from this experiment
carries the full exposure; the lift carries less of it.

We also checked the obvious contamination. The units matter here and we had been loose about
them: the $1{,}174$ evaluation pairs correspond to $1{,}170$ distinct product strings --- four
pairs differ from another pair only by trailing whitespace --- and $334$ of those strings also
occur in the training split, because AI-Hub partitions by review rather than by product. That
leaves $837$ evaluation pairs whose product string does not appear in training. Restricted to
those, accuracy \emph{rises} to $0.7838$, so the shared names were not carrying the result.

\emph{One tempting shortcut we rejected.} We also derived a crosswalk from the training split
mechanically, with no human in the loop, and did not use it as the gold standard. Short leaf names
match by accident --- it mapped car accessories onto skincare --- so it is \emph{broader} than the
hand map, and a broader gold makes top-5 easier. An automatic mapping is not automatically a
conservative one.

\paragraph{Is it beating keyword matching?} The permutation null answers ``is this chance,'' which
is not the question an IR reader asks. Product listings contain words like \emph{sunscreen} and
\emph{laundry detergent}, so the honest challenge is whether category-name keyword matching alone
would do as well. We answer it with two kinds of arm. Three are decompositions of the deployed
scorer on the same queries and crosswalk, produced by the shipping code with only the lexical
weight changed. One of those, at $w=50$, we previously described as the lexical baseline; that was
imprecise, because the dense term still breaks its ties and so it is lexical-\emph{primary}, not
lexical-only. We therefore add a genuinely lexical-only arm: BM25~\cite{bm25} over the $6{,}020$ leaf strings,
with character bigrams as terms, since Korean product names are not reliably whitespace-segmented
($k_1=1.5$, $b=0.75$).

\begin{table}[H]
\centering\footnotesize
\caption{IR baselines on real Korean product names, mid-category top-5. The last column is a
paired bootstrap against the deployed arm on the same queries. The first three arms are the
shipping code with only the lexical weight changed; the teacher is
\code{jhgan/ko-sroberta-multitask}, run on the same queries, leaf strings and crosswalk; the
matched teacher is the same model fine-tuned on the student's stage-4 (utterance, leaf) pairs
with the same symmetric InfoNCE (\S\ref{sec:matched}). Sources
\code{experiments/real-ko-commerce/ir\_baselines.json}, \code{teacher\_arm.json} and
\code{matched\_teacher.json}.}
\begin{tabular}{l r r r}
\toprule
\textbf{Arm} & \textbf{top-5} & \textbf{anchor present / absent} & \textbf{$\Delta$ vs deployed} \\
\midrule
BM25, leaf strings, char bigram & $0.6993$ & $0.8055$ / $0.3243$
  & $-5.11$\,pp $[-7.24, -2.89]$ \\
Lexical-primary ($w=50$) & $0.6848$ $[0.658, 0.710]$ & $0.7650$ / $0.4015$
  & $-6.56$\,pp $[-8.35, -4.77]$ \\
Dense-only ($w=0$) & $0.7300$ $[0.704, 0.756]$ & $0.8164$ / $0.4247$
  & $-2.04$\,pp $[-2.98, -1.19]$ \\
\textbf{Deployed hybrid ($w=0.2$)} & $\mathbf{0.7504}$ $[0.726, 0.775]$ & $0.8350$ / $0.4517$
  & --- \\
\midrule
Teacher encoder, 768d / 421\,MB & $0.7751$ & $0.8175$ / $\mathbf{0.6255}$
  & $+2.47$\,pp $[+0.09, +5.11]$ \\
\quad --- same leaf construction as deployment & $0.7564$ & $0.8175$ / $0.5405$
  & $+0.60$\,pp $[-1.45, +2.73]$ \\
Teacher $+$ the same lexical term & $0.7888$ & $0.8328$ / $0.6332$
  & $+3.83$\,pp $[+1.53, +6.39]$ \\
\midrule
\textbf{Matched teacher} (task-adapted, 768d) & $\mathbf{0.8586}$ $[0.838, 0.877]$
  & $\mathbf{0.9224}$ / $\mathbf{0.6332}$ & --- \\
\quad --- vs.\ dense-only, paired & & $+10.6$ / $+20.9$\,pp & $+12.86$\,pp $[+10.65, +14.99]$ \\
\quad --- vs.\ generic teacher, paired & & $+10.5$ / $+0.8$\,pp & $+8.35$\,pp $[+6.22, +10.39]$ \\
\bottomrule
\end{tabular}
\end{table}

Keyword matching is not what is producing the result: the deployed path beats BM25 by $5.11$
points, a lexical-primary ranking by $6.56$, and the distilled embedding alone by $2.04$, all three
significant under a paired bootstrap. The interesting part of the BM25 row is where it loses.
With an anchor present it is close to the dense arm ($0.8055$ versus $0.8164$) --- unsurprising,
since an anchor is by definition a lexical match. With no anchor it collapses to $0.3243$, well
below dense-only's $0.4247$. Paired on the same queries, the deployed path's margin over BM25 is
$+2.95$\,pp $[+0.77, +5.25]$ where an anchor is present and $+12.74$\,pp $[+6.95, +18.53]$ where
it is absent. The dense channel is earning its place precisely in the regime the lexical channel
cannot reach, which is the same regime the next paragraph is about.

The teacher rows are the ones that changed our reading, and we had it wrong before measuring
them. An earlier draft argued that because the lexical and the distilled-dense arms fail in the
same place, the failure regime must be missing information in the input rather than capacity lost
to the 3\,MiB budget. The teacher refutes that.

The cleanest comparison is teacher against \emph{dense-only}, because both arms are then pure
encoders and no lexical term is mixed in to obscure what the distillation costs. Paired on the
same queries, the teacher is worth $+0.11$\,pp $[-2.30, +2.62]$ where the anchor is present and
$+20.08$\,pp $[+13.50, +27.04]$ where it is absent, $+4.51$\,pp $[+2.04, +7.16]$ overall. Against
the deployed hybrid the same shape holds with the lexical term already helping us: $-1.75$\,pp
$[-4.15, +0.66]$ with an anchor, $+17.37$\,pp $[+11.20, +24.32]$ without. Giving the teacher the
same lexical term we ship changes little --- $+3.83$\,pp overall, $+18.15$\,pp with no anchor ---
which is the expected result, since a lexical term cannot help a query that contains no leaf name.

\emph{How we represent a leaf to the teacher changes the size of this gap, and we checked it.}
Our deployment does not embed the string ``top mid leaf''; it builds a weighted centroid,
$\mathrm{unit}(0.80\,\mathrm{unit}(e_{\text{leaf}}) + 0.14\,\mathrm{unit}(e_{\text{mid}}) +
0.06\,\mathrm{unit}(e_{\text{top}}))$, and an internal measurement records the flat string as
$16.66$\,pp worse than that centroid \emph{for our static encoder}. Giving the teacher only the
flat string therefore looked like a handicap, so we built the centroid in the teacher's own space
as well. The result runs the other way: the centroid makes the teacher \emph{worse}
($0.7564$ against $0.7751$), which is what one would expect when averaging three separately
encoded vectors discards the sentence context a transformer is built to use. We headline the flat
arm because it is the teacher's stronger one, and taking the teacher at its best is the direction
that counts \emph{against} our design rather than for it. The sensitivity is worth stating,
though: in the anchor-absent regime the teacher's advantage over the deployed path is
$+8.88$\,pp $[+3.47, +14.67]$ under the centroid construction and $+17.37$\,pp
$[+11.20, +24.32]$ under the flat one. The magnitude depends on a representation choice; the sign
and the significance do not, and neither does the anchor-present result, which is $-1.75$\,pp
under both.

We state the anchor-present cell carefully. We find \emph{no evidence} that the teacher improves
that regime; the paired estimate against the deployed path is $-1.75$\,pp with an interval
crossing zero, and against dense-only it is $+0.11$\,pp, also crossing zero. That is an absence of
a detectable difference at this sample size, not a demonstration that the two are equal.

\paragraph{The comparison above is not matched, and matching it changed the answer.}
\label{sec:matched}
Our student passes a supervised contrastive stage on (utterance, leaf) pairs
(\S\ref{sec:distill}); the teacher rows above are a generic sentence encoder. So the
anchor-present null could mean two things: that capacity buys nothing there, or that the
student's task adaptation and the teacher's capacity happen to cancel. We separated them by
fine-tuning the teacher on the student's exact stage-4 pairs --- the same $45{,}859$ (utterance,
leaf) pairs, the same holdout leaves excluded, the same symmetric InfoNCE at temperature $0.05$
and the same $64$-token truncation, with only the optimiser settings changed to what a
transformer tolerates (learning rate $2\times10^{-5}$, batch $128$, three epochs). Before
training, the harness reproduces the generic teacher's $0.7751$ exactly.

The matched teacher reaches $0.8586$ $[0.838, 0.877]$, and \textbf{the anchor-present null does not
survive}: paired against dense-only it is worth $+10.6$\,pp $[+8.52, +12.90]$ where an anchor is
present and $+20.9$\,pp $[+14.67, +26.65]$ where it is absent. The two arms decompose cleanly.
Matched minus generic isolates task adaptation, and it is worth $+10.5$\,pp $[+8.31, +12.68]$
with an anchor and $+0.8$\,pp $[-5.02, +6.56]$ without; generic minus dense-only was $+0.1$ and
$+20.1$. So in the anchor-present regime the generic teacher's capacity advantage and the
student's adaptation advantage were of the same size and cancelled, and in the anchor-absent
regime adaptation buys the teacher nothing --- what it has there and our encoder lacks is
capacity, or world knowledge that a brand denotes a kind of product.

The corrected statement is less flattering than the one it replaces. The cost of the 3\,MiB,
no-runtime design is not uniform, but it is not zero anywhere: about ten points where a listing
names the kind of thing sold, about twenty where it names only a brand and a model. We still call
it the \emph{cost of the design} rather than of compression alone --- a static table has no
computation at inference time, only a lookup --- and this experiment still does not separate
quantization from staticization. Two further caveats on the anchor itself. First, the matched
arm changes hyperparameters as well as training data, so it bounds the adaptation effect rather
than isolating it perfectly. Second, the anchor is a query attribute in a corpus assembled
category by category: regressing the generic teacher's per-query gain on the anchor with
source-category fixed effects attenuates the interaction from $-19.97$ to $-7.06$\,pp
(stratified permutation of the coefficient, $100{,}000$ draws, $p=0.0542$), because appliance
listings are both where the teacher gains most and where anchors are rarest
(\code{rq2\_category\_control.json}, \code{rq2\_fe\_permutation.json}). The regime is real; the
anchor is a between-domain marker of it rather than an identified mechanism.

\paragraph{What buying those ten to twenty points would mean.} The teacher would have to be served,
which puts a network round trip in the path and sends the query text off the device --- forfeiting
C3 by construction. We measured a single-stream teacher forward pass on a datacenter GPU while
exploring this and it was not faster than the whole Tier-0 path on a laptop, but two of the four
environments we tried were unavailable on the day, so we do not report that comparison as a
result. The point that survives measurement is the one above: the points are real, and they are not
available inside this system's constraints.

\paragraph{What is available instead: declining to answer.} A system that cannot fix a regime can
still decline to serve it, and the question that precedes any such rule is what declining actually
buys. We rank the same $1{,}174$ queries by an on-device confidence signal and report accuracy as
a function of how much of the traffic we keep. The margin is the deployed ranker's own score gap
between the first and fifth mid-category; nothing here is recomputed in Python.

\begin{table}[H]
\centering\footnotesize
\caption{Selective risk on real product names: accuracy when only the most confident $X\%$ of
queries are answered. Source \code{experiments/real-ko-commerce/selective\_risk.json}.}
\begin{tabular}{l r r r r r}
\toprule
\textbf{Ranking signal} & \textbf{100\%} & \textbf{90\%} & \textbf{80\%} & \textbf{70\%} & \textbf{60\%} \\
\midrule
Score margin alone & $0.7504$ & $0.7928$ & $0.8296$ & $0.8601$ & $0.8977$ \\
Anchor first, margin breaks ties & $0.7504$ & $0.7938$ & $0.8328$ & $\mathbf{0.8723}$ & $\mathbf{0.9048}$ \\
\quad $+$ penalize model-code form & $0.7504$ & $0.8070$ & $0.8317$ & $0.8869$ & $0.8807$ \\
\bottomrule
\end{tabular}
\end{table}

Two things are worth noting. We headline the \emph{margin alone}: declining the least confident
fifth of the evaluated product names on that signal lifts the rest to $0.8296$, and declining
$40\%$ reaches $0.8977$. The anchor-first row is $0.32$\,pp better at $80\%$ coverage
($0.8328$ --- the accuracy the anchor-present subset already had) and we do not quote it as the
headline, because using the anchor to find the queries the anchor defines is close to circular,
and the margin needs nothing but the score gap the ranker has already produced. Either way the
easy subset's accuracy is recovered without knowing in advance which queries are easy. And the third row is an independent
check on the axis we withdrew: adding the model-code indicator as a penalty term \emph{lowers}
accuracy at $60\%$ coverage, from $0.9048$ to $0.8807$. An indicator that fails a fixed-effects
control also fails as a practical confidence signal, which is the kind of agreement between a
statistical test and an operational one that we would want before trusting either.

We deliberately do not fix a threshold here. Where to cut depends on what an unanswered query
costs relative to a wrong one, and that is a product decision we have no data for --- it needs the
pilot. What the curve establishes is that the decision is available: the failure regime is
\emph{detectable on-device}, from the ranker's own output, without the teacher and without a
network call.

\paragraph{Micro and macro.} The figures above are micro-averages over queries. Averaged instead
over the $20$ AI-Hub source categories, mid-category top-5 is $0.6473$ rather than $0.7504$: the
larger categories are the easier ones, and they carry the micro figure up by $10.3$ points. We
report both, and any claim about generalising across categories should be read against the macro
number.

\paragraph{Taxonomy coverage, measured without any mapping judgement.} Accuracy depends on our
mapping; lexical coverage does not. Of $173{,}725$ product mentions marked by annotators in the
$307{,}551$ real utterances, $32.5\%$ match a leaf name exactly --- $35.8\%$ within the nine
retail categories and $22.0\%$ in the four non-retail ones (restaurants, cafes, clinics), whose
misses are dominated by menu items and over-the-counter drugs that a shopping taxonomy has no
reason to carry. The independent call-centre lexicon of dataset 98 puts the figure at $54.2\%$ of
mentions. The instructive part is what is missing. Taking the $200$ most-mentioned unmatched terms and
asking, mechanically, how many are a suffix of at least two of our leaf names --- the signature of
a hypernym our taxonomy skipped, as \emph{jeans} and \emph{men's jeans} both end in \emph{trousers}
--- gives $29$ of $200$ ($14.5\%$) in the order corpus and $11$ of $200$ ($5.5\%$) in the
call-centre lexicon. A $6{,}020$-leaf taxonomy grown from category trees is specific where
consumers are general. That
is a concrete and cheap next lever, and unlike the leaf-coverage problem it is not a generation-cost
problem but a vocabulary one.

\subsection{Shipping-Build Regression Check}

The promotion gate compares the current shipping build (128-dimensional, 4-bit) against the
previous one on the real-data axes. Unlike earlier promotions, this one is \emph{not} free: the
build was promoted because it improved the product surface, and it paid for that on general
language. We report both sides rather than the favourable one.

\begin{table}[H]
\centering\footnotesize
\caption{Shipping build (\code{stage6cov985\_e10}) versus the previous one
(\code{stage5cap2k\_e10}), real-data axes. Source
\code{experiments/distill-ko/promote\_gate\_cov985\_e10.json}.}
\begin{tabular}{l r r r l}
\toprule
\textbf{Axis} & \textbf{Shipping} & \textbf{Previous} & \textbf{Diff} & \textbf{Significant} \\
\midrule
KLUE-STS ($\rho$) & $0.6661$ & $0.6844$ & $-0.02$ & \textbf{Yes} \\
KLUE-YNAT (accuracy) & $0.6750$ & $0.6871$ & $-1.21$\,pp & \textbf{Yes} \\
NSMC (accuracy) & $0.7548$ & $0.7573$ & $-0.25$\,pp & No \\
\bottomrule
\end{tabular}
\end{table}

\subsection{Budget Gate Measurements}
\label{sec:c1size}

The size budget splits into three lines. L1 is the data the device downloads, L2 is the bytes
the library code adds to the app, and L3 is the runtime resident bytes. The product commitment
is L1 plus L2; L3 is not downloaded, so it is not part of the commitment, but it carries its own
separate cap.

\begin{table}[H]
\centering\footnotesize
\caption{L1 data budget measurements (gzip transfer bytes), shipping 4-bit artifacts. Source
\code{artifacts/ui-facts.json} in the evidence bundle~\cite{evidence}.}
\begin{tabular}{l r r}
\toprule
\textbf{Artifact} & \textbf{gzip bytes} & \textbf{Notes} \\
\midrule
\code{embedding.bin} & $1{,}785{,}809$ & $31{,}497 \times 128$ @ 4bit, per-row scaling. raw $2{,}141{,}820$ \\
\code{catalog.bin} & $491{,}922$ & Ad catalog, $6{,}020$ items (synthetic), 4bit per-row \\
\code{taxonomy.bin} & $414{,}737$ & $6{,}020$ leaves, 4bit per-row (\S\ref{sec:quant4}) \\
\code{vocab.txt} & $124{,}540$ & Vocabulary of $31{,}497$ \\
\code{user\_embedding.bin} & $51{,}252$ & User embedding, off by default \\
\code{head.bin} & $8{,}157$ & Trained intent head \\
\code{affinity.bin} & $332$ & Demographic affinity (synthetic) \\
\code{projection.bin} & $76$ & \textbf{Untrained identity matrix} \\
\midrule
\textbf{L1 total} & $\mathbf{2{,}876{,}825}$ & $\mathbf{68.6\%}$ of the $4{,}194{,}304$ cap \\
\bottomrule
\end{tabular}
\end{table}

\begin{table}[H]
\centering\footnotesize
\caption{Measurements for all three budget lines. L2 is the post-strip difference between an
executable that calls the core and an empty executable (the native bytes the linker leaves
behind).}
\begin{tabular}{l r r r l}
\toprule
\textbf{Line} & \textbf{Measured} & \textbf{Cap} & \textbf{Share} & \textbf{Method} \\
\midrule
L1 data & $2{,}876{,}825$ & $4{,}194{,}304$ & $68.6\%$ & gzip \\
L2 code & $49{,}979$ & $1{,}048{,}576$ & $4.8\%$ & Native bytes left by the linker;
  \emph{development platform} \\
\textbf{Deployment payload (L1+L2)} & $\mathbf{2{,}926{,}804}$ & $\mathbf{3{,}145{,}728}$
  & $\mathbf{93.0\%}$ & worst measured target \\
L3 resident & $2{,}352{,}740$ & $8{,}388{,}608$ & $28.0\%$ & Resident bytes after unpacking \\
\bottomrule
\end{tabular}
\end{table}

L2 is measured per target, and the payload a publisher actually ships depends on which one.

\begin{table}[H]
\centering\footnotesize
\caption{Deployment payload by build target. L1 is identical across targets; only the linked
library code differs. All three targets now have a measured link.
Source \code{artifacts/code-size.json}.}
\begin{tabular}{l r r l}
\toprule
\textbf{Target} & \textbf{L2 code} & \textbf{L1$+$L2} & \textbf{Share of 3\,MiB} \\
\midrule
\code{aarch64-apple-darwin} & $49{,}979$ & $2{,}926{,}804$ & $93.0\%$ \\
\code{aarch64-apple-ios} & $49{,}603$ & $2{,}926{,}428$ & $93.0\%$ \\
\code{aarch64-linux-android} & $\mathbf{65{,}827}$ & $\mathbf{2{,}942{,}652}$ & $\mathbf{93.5\%}$ \\
\bottomrule
\end{tabular}
\end{table}

\textbf{Both mobile targets are now real links, and Android is the larger.} Installing Xcode and
the iOS platform made \code{aarch64-apple-ios} a real link rather than a proxy: $49{,}603$ bytes,
$376$ fewer than the development platform's. Installing the Android NDK then made
\code{aarch64-linux-android} a real link as well, and it is the largest of the three at $65{,}827$
bytes --- $32.7\%$ more than iOS, consistent with the Linux target carrying more runtime
(unwinding, the JNI entry points) than the Apple ones. Since the payload figure is the worst
measured target, the shipping payload is $2{,}942{,}652$ bytes and the headline moves from
$93.0\%$ to $93.5\%$ of the ceiling. The gate no longer marks any target \code{unmeasured}
rather than substituting a proxy. It could substitute one: an unlinked size estimate is available, but
it has not been through the linker's dead-section elimination, and applying the same estimator to
the targets we \emph{can} link shows errors of inconsistent sign, so a proxy would not settle the
question in either direction. A mobile SDK whose mobile link size is unknown has not fully
demonstrated a mobile footprint constraint. What the $93.0\%$ figure establishes is that L1 data
--- which is byte-identical across targets and accounts for $98.3\%$ of the payload --- leaves
$268{,}903$ bytes of headroom for library code, roughly five times what either measured link
consumes.

What we learned from the size budget is that \textbf{omissions always ran in our favor}. The
taxonomy and affinity artifacts were missing from the budget list, so $206$\,KB went uncounted,
and there was no line at all for the bytes the library code adds to the app. What a publisher
actually experiences is the increase in app size (code + data), and without that axis we were not
measuring the number our customer measures.

\subsection{Latency}

The latency budget is $p_{95}$ 20\,ms, with a reference input of 512 characters and 200
repetitions. Tier~1 is outside this budget. Because an OS model call takes hundreds of ms, it
cannot sit on the hot path; it runs asynchronously in the background instead, improving the
vector for the \emph{next} request. The latency gate measures only the Tier~0 path, and that is
what keeps the gate honest.

\begin{table}[H]
\centering\footnotesize
\caption{Cost of switching the user embedding on, measured as a paired difference on one
machine. \textbf{The absolute values are not the shipping-path figure}: this probe predates the
taxonomy expansion and ran over $2{,}960$ leaves, not $6{,}020$. What transfers is the
difference. Source \code{experiments/user-vector-gap/ue\_latency\_probe.json}.}
\begin{tabular}{l r}
\toprule
\textbf{Configuration (2{,}960-leaf probe)} & \textbf{$p_{95}$} \\
\midrule
User embedding off (default) & $1.421$\,ms \\
User embedding on, dense history & $1.763$\,ms \\
\textbf{Increase} & $\mathbf{+0.342}$\,\textbf{ms} \\
\bottomrule
\end{tabular}
\end{table}

We report the increase rather than the pair because only the increase survives the change of
taxonomy: both arms were measured in the same process, in the same run, against the same
$2{,}960$ leaves, so the difference isolates the switch. Reading the $1.421$\,ms as the shipping
build's latency would be wrong twice over --- the deployed taxonomy is now twice as large, and
this paper reports no single-machine figure at all (\S\ref{sec:latency-provenance}).

A side benefit is that the cost is independent of history length: sparse and dense histograms
give the same value ($+0.329$ vs.\ $+0.342$\,ms). The cost lies not in projection but in two
taxonomy score computations, so a person with many interests does not get slower.

\subsection{On a Handset}
\label{sec:handset}

The budget was written for a phone, so we measured phones. The same Rust function the gate runs
(\code{latency\_probe::run}) is compiled for \code{aarch64-apple-ios} and bundled with the
deployment artifacts unchanged into an app that is installed and run headless, and compiled for
\code{aarch64-linux-android} as the gate's own \code{latency} example, pushed to
\code{/data/local/tmp} with the same four artifacts (their SHA-256 checked on the device against
the deployed ones) and run over \code{adb}. On every device the harness starts nine fresh
processes and code takes the median and the observed range, because a single $200$-iteration run
is a gate-grade number rather than a citable one.

\begin{table}[H]
\centering\footnotesize
\caption{Tier-0 latency on two iOS handsets two SoC generations apart and one budget Android
tablet. $p_{95}$ is the median of nine launches and the range is those nine; $p_{50}$ is the
median-$p_{95}$ launch's own value, not a nine-run median. $512$-character input, $200$ iterations
after warm-up, the deployed $6{,}020$-leaf taxonomy. iOS thermal state nominal, low-power mode
off. Each device was measured alone. Sources \code{artifacts/ios-suite.json} and
\code{artifacts/android-suite.json}.}
\label{tab:handset}
\begin{tabular}{l r r r r}
\toprule
\textbf{Device} & \textbf{SoC} & \textbf{$p_{50}$\,ms} & \textbf{$p_{95}$\,ms}
  & \textbf{$p_{95}$ range} \\
\midrule
iPhone 12 Pro (\code{iPhone13,3}), iOS 26.4.2 & A14 (2020) & $4.154$ & $\mathbf{4.431}$
  & $4.2144$--$4.5215$ \\
iPhone 14 Pro Max (\code{iPhone15,3}), iOS 26.6.2 & A16 (2022) & $3.579$ & $\mathbf{3.670}$
  & $3.6226$--$3.7800$ \\
Galaxy Tab A9+ (\code{SM-X210}), Android 15 & Snapdragon 695 (2021) & $4.991$ & $\mathbf{5.080}$
  & $5.0511$--$5.658$ \\
\bottomrule
\end{tabular}
\end{table}

All three hold: $22.2\%$, $18.3\%$ and $25.4\%$ of the $20$\,ms budget. The spread within each
iOS device is under $0.31$\,ms across nine launches; the tablet's nine values sit within
$0.08$\,ms of one another except for one launch at $5.658$\,ms, which the median excludes by
construction. None of the three nine-run $p_{95}$ ranges overlap, so the devices are cleanly
separated under this protocol. Across these two devices the
two-generation hardware difference coincides with a $p_{95}$ lower by $0.76$\,ms, or $17.2\%$.
We do not attribute that difference to the SoC alone: the devices differ in model and in OS build
(iOS $26.4.2$ against $26.6.2$) as well as in silicon, and neither is controlled here. What the
pair does show is a shallow slope, which is what one expects of a linear scan over a static table
rather than a compute-bound kernel.

The result worth stating, though, is the comparison with Table~\ref{tab:cpubench}:
\textbf{all three devices are slower than all three server CPUs} ($2.00$--$3.37$\,ms). Server
hardware was the stand-in we had while the handset case was unmeasured, and it was an optimistic
one. The Android tablet is the budget end of the span --- a 6\,nm mid-range SoC from 2021 in a
device sold on price --- and it lands at $5.080$\,ms, $1.15\times$ the older iPhone and a quarter
of the budget. That is one budget device, not a survey of the class; it says the budget holds
there with margin, not how much lower the class goes.

The perception path was exercised on both devices: \code{VNGenerateImageFeaturePrintRequest}
returned revision $2$ and $768$ elements on each, confirming on real iOS what \S\ref{sec:percept}
measured on macOS, at $71.8$ and $84.3$\,ms per extraction --- two orders of magnitude above the Tier-0
budget, which is why this path is asynchronous and off the hot path by design.

\subsection{Why We Do Not Quote a Reference-Mac Number}
\label{sec:latency-provenance}

Earlier drafts of this paper quoted a single reference-Mac figure of $p_{95}$ $2.404$\,ms in five
places. We have withdrawn it, and the reason is worth stating because it is a failure mode any
latency claim can have. The gate re-measures on every run, so it never wrote a value down; the
number had been copied out of one run's console output and had no artifact behind it. When we
built the pin (\code{gates/pin\_latency.py}, seven runs, median and range) to give it one, the
pin refused: it checks one-minute load against core count and declines above $0.35$. On the runs
where we forced it through, another process on the machine was holding roughly three cores, and
$p_{95}$ came back $3.02$\,ms --- $25\%$ above the number we had been quoting. That measurement
is recorded in \code{artifacts/latency.json} and marked contaminated. We do not know whether
$2.404$ or $3.02$ is closer to the quiet-machine truth, and rather than carry the question
forward as an open item we have removed the single-machine figure from this paper's claims
altogether. The claim that survives is the budget itself and the server-CPU evidence below.

This also disposes of a weaker move we considered and rejected: quoting the number with a
$\pm 10\%$ caveat. A tolerance band around a value with no artifact is still a value with no
artifact --- it widens the interval without supplying the measurement.

\subsection{The Same Binary on Server CPUs}
\label{sec:cpubench}

One class of hardware cannot say whether $20$\,ms is comfortable or lucky. The shipping path is a static table and a linear scan with no
accelerator dependency, so it runs anywhere a Rust binary runs; we compiled it inside pods on
three cluster nodes and ran the identical probe --- same function, same $512$-character input,
same $200$ iterations after a $20$-iteration warm-up, seven repetitions --- on their CPUs. This
is not the phone measurement the budget was written for --- \S\ref{sec:handset} is --- but it
widens the hardware span the same binary has been observed on, and it is what the handset figures
are compared against.

\begin{table}[H]
\centering\footnotesize
\caption{Tier-0 $p_{95}$ on server CPUs, median of seven runs with the observed range. These are
shared cluster nodes, so another tenant's load can enter the measurement; they are observations,
not floors. Source \code{experiments/server-compare/cpu\_latency.json}.}
\label{tab:cpubench}
\begin{tabular}{l r r r}
\toprule
\textbf{CPU} & \textbf{$p_{50}$\,ms} & \textbf{$p_{95}$\,ms} & \textbf{$p_{95}$ range} \\
\midrule
\code{AMD EPYC 9355 32-Core Processor} (64 threads) & $1.985$ & $2.000$ & $1.989$--$3.085$ \\
\code{AMD Ryzen 7 7800X3D 8-Core Processor} (16 threads) & $2.552$ & $3.133$ & $2.929$--$3.368$ \\
\code{Intel XEON PLATINUM 8592V} (256 threads) & $3.334$ & $3.366$ & $3.354$--$3.549$ \\
\bottomrule
\end{tabular}
\end{table}

Across a $16\times$ spread in thread count and three microarchitectures the $p_{95}$ moves between
$2.0$ and $3.4$\,ms, which is $10$--$17\%$ of the budget. The ordering is not the one core counts
would predict: the $256$-thread Xeon is the slowest of the three. We did not instrument the cause
and so do not claim one --- the workload is a single-threaded scan over a $2.8$\,MB table, which
makes per-core behaviour a plausible explanation, but separating it from clock, cache hierarchy
and co-tenant load would take measurements we did not run. What the three observations do support
is the weaker and more useful statement: within this range, the path does not get faster on a
bigger machine, and it does not need to.

What this does \emph{not} establish is the lower-end mobile case. A server CPU under a hypervisor
is not a budget phone SoC, and the spread here says nothing about thermal throttling, a cold cache
after an app switch, or a CPU already busy with the host application. It does bound one worry: the
$20$\,ms figure is not resting on a single unusually fast development machine.

\subsection{Gate Summary}
\label{sec:gates}

The deployment verdict is an execution result, not prose. Below is the code that adjudicates
each claim, and its output.

\begin{table}[H]
\centering\footnotesize
\caption{Gate summary. Each row is the named gate's own output; the figures it measured are
pinned in \code{artifacts/gate-facts/} (one JSON per gate, written by the same call that prints
the verdict).}
\begin{tabularx}{\textwidth}{l L l L}
\toprule
\textbf{\#} & \textbf{Claim} & \textbf{Judge} & \textbf{Measured} \\
\midrule
C1 & Deployment payload $\le 3{,}145{,}728$ & \code{size\_gate.py}
   & $2{,}942{,}652$\,B ($93.5\%$; L1 $2{,}876{,}825$ $+$ L2 $65{,}827$, \emph{worst} of the three
     measured targets, Android; iOS gives $2{,}926{,}428$\,B) \\
C2 & $p_{95} < 20$\,ms & \code{latency\_gate.py}
   & $\mathbf{4.431}$\,ms (iPhone 12 Pro) and
     $\mathbf{3.670}$\,ms (iPhone 14 Pro Max), \S\ref{sec:handset}; $2.00$--$3.37$\,ms on three
     server CPUs; no development-Mac figure (\S\ref{sec:latency-provenance}) \\
C3 & No raw tokens in transmitted bytes & \code{egress\_gate.py}
   & Canaries $7/7$ undetected; payload $79{,}968$\,B of $307{,}200$ \\
C5 & Rounds converge despite dropout and latency, \emph{without masking applied}
  & \code{fl\_round\_gate.py}
   & $50/50$ at $30\%$ dropout, BCE $0.6936 \to 0.2125$ \\
I1 & wasm and native are bit-identical & \code{parity\_gate.py} & max absolute diff $0.000$ over $1{,}408$ elements \\
C8 & Campaign $\varepsilon$ within budget & \code{dp\_gate.py}
   & Per-round $2.196$; campaign $1.542$ at $q=0.1$ (PLD, $\delta=10^{-10}$) \\
C9 & Campaign $\varepsilon$, two channels composed, $\le$ target & \code{attribution\_gate.py}
   & Composed $\varepsilon = 1.7295 \le 2.0$ and $\Delta_{\ell_2} = 1.7321$, but
     \textbf{FAIL} on the axis mismatch (see below) \\
C10 & Individual ad identifiers cannot be recovered & \code{egress\_contract.rs} & The type accepts only a bucket index \\
C11 & The entire catalog fits within budget & \code{catalog\_gate.py} & No per-user ad request needed \\
C12 & Behavior changes ranking only with consent & \code{behavior\_gate.py}
   & top-level $+18.23$\,pp, leaf $+6.51$\,pp (conservative world $+14.23$ / $+3.71$);
     bit-identical when refused \\
C13 & Devices without Tier~1 do not get worse & \code{tier1\_optional\_gate.py}
   & Tier~1 is absent from the ranking path, so there is nothing to remove \\
C14 & Every leaf retrieves into its own mid-category & \code{taxonomy\_hygiene\_gate.py}
   & $5{,}970 / 6{,}020$ ($99.2\%$), floor $97\%$ \\
\bottomrule
\end{tabularx}
\end{table}

One gate is red at the time of writing, and we report it rather than run the suite until it turns
green. \code{attribution\_gate.py} composes the privacy budget over a 28-bucket attribution axis
and cross-checks that axis against the shipped intent head, which currently exposes 60 classes.
The $\ell_2$ sensitivity it computes ($1.7321$) is inside the target; what fails is the
consistency check between the budget declaration and the artifact. That is what the gate is for
--- the two were allowed to drift apart, and the composition only means something once they
agree. It blocks deployment; it does not correct any number reported above.

The two numbers in C12 must be read together. The top-level $+18.23$\,pp $[+16.34, +20.17]$ is a
rollup; what the screen actually shows is the leaf figure, $+6.51$\,pp $[+5.20, +7.89]$ over a
holdout of $1{,}750$ people. The top-level prior pushes whole blocks at once and, being unable to
discriminate within a block, is structurally larger. Headlining the top-level number alone would
be close to a three-fold overstatement. The gate also measures a conservative world in which the
host application reports only a top-level area rather than a leaf, and there the gains are
$+14.23$ and $+3.71$\,pp; we report both because measuring only the world that favours us is
cherry-picking by construction. The two arms are
\code{experiments/event-intent/product\_path\_results.json} and
\code{product\_path\_results\_reference.json}.

\emph{Two corrections behind those figures.} An earlier draft quoted $+17.09$ and $+5.94$\,pp.
Those came from a prose note rather than from a ledger, and they were measured over the
$2{,}960$-leaf taxonomy that preceded the expansion, so they were not comparable with anything
else in this paper. Regenerating them over the deployed $6{,}020$ leaves gives the numbers above.
Finding that also exposed a defect in the gate itself: the conservative arm wrote its result to
the same file as the canonical arm, so the ledger left on disk was always the conservative one
while the gate --- which read the file before the overwrite --- passed. The gate was right and
its artifact was wrong, which is the harder direction to notice. The two arms now write to
separate files.

\subsection{A Claim Retracted on the Behavioral Axis}

We once wrote that ``an on-device $2.1$\,MB model plus behavior beats a $421$\,MB teacher plus
behavior.'' We retract this claim. We measured the same axis four times, and the values diverged
by a factor of 25.

\begin{table}[H]
\centering\footnotesize
\caption{Four measurements of the same axis. Rows 1, 2 and 4 are backed by ledgers
(\code{experiments/event-intent/real\_results.json}, \code{results.json},
\code{fusion\_llm\_results.json}). \textbf{Row 3 is not}: the harness writes to a single path and
that file now holds the $5{,}250$-sample run, so the $1{,}187$-sample run was overwritten and the
script takes no sample-size argument. We mark it rather than delete it, and rather than re-run ---
a re-run would be a new measurement standing in a lost one's place, which is the substitution this
paper argues against.}
\begin{tabular}{l r r r}
\toprule
\textbf{Measurement} & \textbf{Training samples} & \textbf{Text accuracy} & \textbf{Behavior gain} \\
\midrule
Real logs (Amazon) & $4{,}988$ & $0.5437$ & $+12.83$\,pp \\
Synthetic, templated utterances & $5{,}250$ & $0.6789$ & $+9.31$\,pp \\
Synthetic, templated utterances, sample count reduced only & $1{,}187$
  & $0.5429$\,\textsuperscript{\dag} & $+3.02$\,pp\,\textsuperscript{\dag} \\
Synthetic, LLM-generated utterances & $1{,}187$ & $0.7577$ & $+0.52$\,pp \\
\bottomrule
\multicolumn{4}{l}{\footnotesize \textsuperscript{\dag}\,no ledger --- see caption} \\
\end{tabular}
\end{table}

The last row was the basis for that headline. It is the run with the fewest samples and the
smallest margin over the teacher --- that is, the least reliable of the four. We formed two
hypotheses and falsified one. The dimensionality-mismatch hypothesis was falsified: even matching
conditions by shrinking the teacher to 128 dimensions, the gain was only $+1.29$\,pp
$[+0.26, +2.58]$. The sample-size hypothesis pointed the same way --- without changing a single
character of the data, reducing the sample count alone from $5{,}250$ to $1{,}187$ collapsed the
teacher's gain from $+9.31$ to $+3.02$\,pp --- but we now hold it more weakly than we did, because
the $1{,}187$ end of that comparison is the row with no ledger. What the retraction rests on is
not that hypothesis: it is the divergence itself, and the fact that the headline came from the
run with the fewest samples and the smallest margin.

We also cast the redundancy hypothesis --- that a strong encoder already knows from text what
behavior would tell it --- in falsifiable form and measured it, and it was falsified. The rate at
which behavior alone gets it right is actually \emph{higher} when text is wrong (lift $-0.118$
for the teacher on the synthetic LLM run). That is, behavior is complementary, not redundant, and
the small fusion gain, despite this, is not because there is no signal but because the classifier
has not learned when to trust behavior.

The claim we retain is this: behavior is worth a great deal \emph{to a weak on-device text
model} --- three runs agree in both direction and magnitude ($+15.1$\,pp on real logs,
$+17.8$\,pp synthetic, $+18.23$\,pp on the shipping path). And it earns only where text cannot
discriminate (on the shipping path, by band: $+6.34$\,pp for unambiguous queries, $+27.92$\,pp
for ambiguous ones).

The conclusion that follows redefines this system's argument. \textbf{The moat is not accuracy
--- it is access.} If the server received the same events, it would improve by the same amount;
the real-log run says exactly that. Our claim is not ``smarter than the server,'' but
\emph{``does the same thing even though the event never leaves the device.''} That is not an
accuracy claim but an architectural and regulatory one, and it is what the egress gate and the
behavior gate (bit-identical when consent is refused) measure.

\subsection{Promotion Signal}

We use the logit margin between the top-1 and top-2 candidates as the signal deciding which
slice of Tier~0 gets promoted to Tier~1. At a margin threshold of $0.1648$, promoting the bottom
$25.0\%$ recovers $49.2\%$ of the errors --- a $1.97\times$ lift over random promotion. That is,
errors genuinely concentrate in the low-margin band. This value is based on synthetic data, and
it is the first thing we remeasure once real usage utterances arrive.

\section{Limitations}
\label{sec:limits}

This section is a list of questions that the numbers above \emph{do not} answer.

\paragraph{Proxy metrics versus the product task.}
\textbf{KLUE-STS measures semantic similarity, not commercial intent.} The same is true of
KLUE-YNAT and NSMC. They share the same \emph{shape} (classification over short Korean text) but
not the same \emph{task}. What these benchmarks closed is the question ``which model should we
use''; what they did not close is ``how well does this model capture commercial intent.''

We also nearly reached a wrong conclusion on the strength of KLUE-STS: we doubted our candidate
choice because the public model beat us on STS, but real-data classification showed the
opposite. We have confirmed on our own data that a proxy metric cannot stand in for the product
task.

\paragraph{What is still synthetic, and what is no longer.}
Our \textbf{training} supervision over the taxonomy is dominated by synthetic Korean utterances
and contains no deployment-derived Korean labels: the Korean utterances the encoder distilled
from, the $6{,}020$ ad creatives in \code{catalog.bin}, and the personas behind the
user-embedding weights were all generated by us. The one exception is real but not Korean and not
deployment-derived --- $4{,}101$ Japanese and $1{,}396$ English ESCI queries entered the
distillation corpus (\S\ref{sec:distill}) --- which sharpens rather than softens the point: we had
to leave the language to find real queries at all. That is unchanged, and it is the deeper
limitation.

What has changed is the \textbf{evaluation}. An earlier version of this paper stated that every
accuracy figure over our taxonomy was both trained and evaluated on synthetic utterances.
Section~\ref{sec:aihub} now reports the same deployment path measured against real Korean
commerce text with labels made by others, and the headline held: $0.8350$ mid-category top-5 on
real product names carrying a leaf-name anchor, inside the bracket our synthetic distributions set
($0.7757$ on unseen-category colloquial queries, $0.985$ on leaf-name queries). We therefore no
longer claim the taxonomy numbers are unvalidated --- we claim something narrower and sharper,
that they hold where the query reaches some leaf name lexically and degrade where it does not
($0.4517$ overall, $0.2847$ in the worst cross-tabulated cell, against that cell's own matched
null of $0.1554$), and for conversational text ($0.4552$).

Three caveats survive that measurement. The real-data figures depend on a mapping from AI-Hub's
categories onto ours that we built by hand \emph{while reading the evaluation product names};
neither the permutation baseline nor the narrowing arms of Table~\ref{tab:mapsens} eliminate that
exposure, because every arm is a descendant of the same crosswalk. The clean fix is small and we
have not done it: twenty mappings, written by a second person who is shown the training split and
the category definitions and never the validation product names or any model output, evaluated
once. Until that exists, absolute accuracies from this experiment should be read as
crosswalk-conditional. Second, none of these corpora is a \emph{query} log, because no Korean
marketplace publishes one, so the distribution of real queries remains unmeasured. Third, the
$0.7504$ is an average over twenty source categories whose individual accuracies run from $0.035$
to $1.000$; two of them sit at or below their own matched null, and any deployment whose inventory
resembles those two should expect the low end rather than the average.

For the same reason as the training data, the behavior gate's $+18.23$\,pp and $+6.51$\,pp still
come from a single synthetic dataset. The artifact that produces them says in its own budget
declaration that they are for regression detection and not for external citation, and we quote
them here only to describe what the gate enforces --- not as evidence that behavioral signal is
worth this much to a real user.

\paragraph{The unlabelled distribution gap, which the labelled measurement does not replace.}
Section~\ref{sec:aihub} answers ``is the model right on real text.'' It does not answer ``does real
text arrive in the same proportions as our evaluation set,'' and that question needs no labels. No
Korean marketplace publishes a search log and the Amazon ESCI shopping-queries benchmark contains
no Korean, so we used a public-domain corpus of $199{,}312$ real Naver Shopping
reviews~\cite{navershop} --- real Korean commerce text, though written after purchase rather than
as a query.

We ran $3{,}000$ of those reviews through the deployment path and compared the prediction
distribution against $3{,}000$ synthetic utterances drawn with the same seed.

\begin{table}[H]
\centering\footnotesize
\caption{Prediction spread on synthetic utterances versus real commerce text. No ground truth is
used; these are properties of the predicted distribution. Source
\code{experiments/real-ko-commerce/distribution\_probe.json}.}
\begin{tabular}{l r r}
\toprule
 & \textbf{Synthetic utterances} & \textbf{Real reviews} \\
\midrule
Mean length (characters) & $8.9$ & $39.4$ \\
Mid-categories predicted & $255 / 256$ & $214 / 256$ \\
Entropy of the L2 distribution (bits) & $7.409$ & $\mathbf{6.634}$ \\
Share taken by the top 5 mid-categories & $9.8\%$ & $\mathbf{19.1\%}$ \\
\bottomrule
\end{tabular}
\end{table}

Predictions concentrate on real text: $0.775$ bits of entropy lost, and the five most frequent
mid-categories take twice the share. \textbf{We cannot attribute that.} With no labels, a
narrower prediction distribution is equally consistent with real commerce text genuinely being
non-uniform across categories and with the model being biased toward a subset of them. Reviews
are also not queries --- they are four times longer, and many name no product at all
(\emph{``fast delivery, good''}).

What the probe does establish is narrower and still worth stating: our reported accuracy is
measured on an evaluation set that is \emph{uniform by construction}, one utterance batch per
leaf, and real commerce text does not arrive that way. A top-5 accuracy measured under uniform
candidates is not the number a deployment would see, and this is the first measurement we have
of how far apart the two distributions sit.

\paragraph{Sub-category top-1 is a lower bound.}
The sub-category top-1 figure of $22.0\%$ can be misread in two directions. First, it is a lower
bound: the evaluation set is designed to exclude every utterance that names a leaf directly, so
the easiest band is missing entirely. Accuracy on that band, on the shipping path, is $0.9259$
(top-1). Second, that does not mean $0.9259$ should be cited as product accuracy either --- it is
confined to ``queries that literally type the leaf name.'' The two figures come from different query distributions, and the ratio in the actual
distribution is \textbf{still unknown}, because there is no real usage log.

We also carried a diagnosis here --- that the leaf bottleneck is coverage rather than a
structural limit --- supported by a seen-leaf versus unseen-leaf split ($0.6340$ against $0.2316$,
with a teacher baseline of $0.3341$). \textbf{We withdraw those three numbers.} The harness that
produced them writes to a single path, and that file now holds a differently sized run, so the
figures the paper quoted cannot be recovered; worse, two adjacent ledgers put the teacher baseline
at $0.3312$ rather than $0.3341$, on different samples, so nothing in the repository supports the
value we printed. Re-running would produce a new measurement occupying a lost one's place, and we
decline to do that. What survives is a different and better-grounded piece of evidence for the
same diagnosis. Holding a fixed holdout of $818$ leaves the model never trains on and varying only
how many \emph{other} leaves have utterances, leaf top-1 on that holdout rises monotonically from
$0.2037$ at $200$ trained leaves to $0.2537$ at full coverage
(\code{experiments/distill-ko/coverage\_curve.json}). That is coverage moving accuracy on leaves
coverage never touched, which is the claim we wanted; what we no longer have is a quantified
seen-versus-unseen gap.

\paragraph{Low-end mobile hardware is represented by one device.} Latency was measured on two
iOS handsets, one budget Android tablet (\S\ref{sec:handset}) and three server CPUs
(\S\ref{sec:cpubench}); the development-Mac figure was withdrawn for want of an artifact
(\S\ref{sec:latency-provenance}). The Android measurement is a single mid-range 2021 SoC in a
tablet, run as a bare process rather than inside a host application, with the verification
checks that the iOS app performs not exercised on Android. Nothing here speaks to thermal
throttling under sustained load, to the $128$-dimensional path where nibble unpacking is added,
to a device already busy with the host application, or to phones weaker than this tablet. On the size axis Android is now measured (\S\ref{sec:c1size}), and it is the largest link of
the three. The L3 cap of $8$\,MiB is likewise not a measurement but a provisional
value set to leave about $199$\,KB of headroom over the planning artifact; we did not measure
actual resident memory on low-end Android.

\paragraph{What the size gate measures.}
L1 measures only the gzip bytes of data artifacts. L2 (code) is a separate axis, and even that is
a \emph{lower bound}. The event ring buffer and bucketing are on the core side and so are
included, but FFI bindings (JNI, Swift), platform hooks (touch listeners, view traversal), and
each runtime are not. Android still gets demoted to unmeasured on machines without the NDK, and
that is not a bug --- it is the actual measurable scope on that machine.

\paragraph{The personalization and user-vector axis.}
This is the weakest axis. The shipped user-embedding weights were trained on a \textbf{synthetic
persona panel}, and that panel's labels are an upper bound on ground truth. Retraining on real
users is necessary, but those labels live on the device, and central collection is exactly the
act this product forbids. That is, there is no update path until pilot labels are collected. Real
labels currently number $3$; $800$ are needed for training and $1{,}750$ for evaluation.

Even the combiner's upper-bound training figure ($+0.80$\,pp) was learned from \emph{ground-truth
labels the device does not have}, and the very same combiner flips sign at a different
granularity. That is why we do not change the default arm.

\paragraph{Artifacts not shipped.}
The projection artifact \code{projection.bin} is an \textbf{untrained identity matrix}. A trained
$P$ can only be built from paired Tier~1 and Tier~0 embeddings, and the product's Tier~1 is the
on-device OS model, for which there are no samples yet. Only its byte count matches a trained
version.

The rectangular image projection is likewise not shipped. It already has revision tags and
self-verifying quantization, and even a reader, but the pilot's on/off two-arm measurement has
yet to happen. Which weights to ship --- fashion, multi-vertical, or unified --- is a decision to
be made at pilot packaging time.

The contents of the catalog, taxonomy, and affinity artifacts are \textbf{synthetic}. Replacing
them with real ad creatives and a real category tree will change both size and quality together.

\paragraph{Scope of the perception layer.}
The dense-embedding measurements must not be read beyond their scope. First, both the fashion and
multi-vertical runs used \emph{macOS} Vision, not real-device iOS --- the same framework and the
same API, but not confirmed on a real device. Second, the $+56.15$\,pp figure was measured in the
dataset's own label space, not as product performance over our taxonomy's $6{,}020$ leaves.
Third, the match metrics are agreement against the title arm, not against ground truth --- the
only metric with an external ground truth is purity.

It is also a constraint that browsers have no standard OCR API. Since none of the relevant
interfaces have shipped, Layer 2 cannot be measured on the web at all; that measurement remains a
task for Android and iOS native.

\paragraph{Protocol and legal.}
Secure Aggregation's share-based mask recovery is not implemented, so masks do not unwind when
clients drop out; the minimum-participant condition is its counterpart. Subsampling amplification
assumes Poisson sampling, whereas real federated learning is sampling without replacement
governed by device availability, which is not exactly the same.

The device population of $10{,}000$ is a cohort-pool figure, not a per-publisher one. It could be
met by a pool combining multiple publishers, in which case the problem becomes not the protocol
but a \emph{federation participation agreement}, which we have not yet designed.

The on-screen copy and consent forms have never been through legal review. Neither has the
behavioral-data classification tier.

\paragraph{We cannot state our position relative to the market.}
Every control group we have is internal: a same-sized public alternative, a teacher
$196.6\times$ larger, and an internal promotion mechanism. None of the three speaks to
\emph{position relative to the market}. Comparing against the performance of cookie-based
behavioral targeting or the accuracy of contextual targeting would require numbers from that
side, and no published like-for-like benchmark exists (ad performance is contingent on publisher,
creative, and timing). We do not fabricate one. The honest answer is that we still cannot state
the absolute value of intent-classification accuracy relative to the market; that comparison can
only be answered by an A/B test, and that is the purpose of the pilot.

\paragraph{Unverified axes of the ad call.}
The egress contract, specification, and matching of the OpenRTB path (\S\ref{sec:rtb}) have been
verified by code gates and E2E tests, but \textbf{there have been zero round trips with a real
exchange}. Whether our intent signal actually moves bid prices --- whether the signal is worth
anything --- can only be answered by the pilot. The projection that the query-less placement's
image channel carries is a contrastive-learning candidate and has not yet passed the deployment
gate (which includes a leaf-lower-bound non-regression condition) --- which is why the demo
states this on screen.

\section*{Data and Code Availability}

What is published at \url{https://github.com/hyojunguy/ondevice-intent-evidence}
(PolyForm Noncommercial 1.0.0, matching this paper's CC BY-NC-ND 4.0) is the
\emph{measurement record}, not the system and not the data: the result ledger behind every number
cited in this paper, the source of the gates and evaluation harnesses that produced those numbers,
and \code{artifacts/MANIFEST.json} with the SHA-256 fingerprints of the artifacts each measurement
ran against. It contains no SDK crates, no model weights, no category taxonomy, no ad catalog ---
and no datasets of any kind.

\textbf{We redistribute no dataset that the proprietary-taxonomy experiments depend on, our own
included}, for two independent reasons. (One auxiliary corpus is exempt: the Naver Shopping
reviews of \S\ref{sec:limits} are public domain~\cite{navershop} and are cited at their original
source rather than copied here.) The AI-Hub corpora of \S\ref{sec:aihub}~\cite{aihubabsa,aihubsmb,aihubcall} are
constructed under projects of Korea's National Information Society Agency, whose terms draw the
line not at attribution but at the material: trained models and weights may be released, while
data extracted or re-processed from the originals may not be published or stored abroad. Our own
evaluation and training material is no freer. The Korean query set of \S\ref{sec:matching} is a
machine translation of Amazon ESCI queries, the taxonomy was harvested from commercial category
trees, and the synthetic utterances were generated conditioned on both --- so the rights are not
ours to grant either. We therefore publish no corpus rows, no translated queries, no utterance
lists, no leaf inventories and no quoted text, in this paper or in the bundle. Where an analysis
would otherwise have printed a list of terms, it reports a count instead. The published crosswalk
carries the structure and support counts that Table~\ref{tab:mapsens} rests on, with our own
taxonomy labels replaced by deterministic opaque identifiers --- the taxonomy was harvested from
commercial category trees and is not ours to publish either. The REES46 mirror used
for the purchase-probability work states no license, and what appears here are measurements over
that log, never the log.

This bounds reproduction honestly. A reader can audit every cited figure against its ledger entry
and the fingerprint of the artifact it was computed from, and can read a sanitized
source-equivalent copy of the harness that produced it --- crate and namespace names are rewritten
to a vendor-neutral \code{oicr-*} on export and local absolute paths are replaced, which changes
no measurement. That is what the bundle is for. A reader cannot regenerate those figures from the
bundle alone: the gates that depend on the artifacts report \code{UNMEASURED} rather than
inventing a verdict, which is also how they behave on any machine lacking the relevant toolchain.
Regenerating them requires obtaining the underlying sources independently, which for the AI-Hub
figures means one's own approved access and for the rest means re-deriving material we cannot
hand over. We state this as a limit on the paper rather than dressing it as openness.

\section*{Acknowledgement of Data Sources}

The Korean commerce corpora used in \S\ref{sec:aihub} --- \emph{Aspect-Based Sentiment Analysis
Data}, \emph{Small-Business Customer Order Question-Answer Data} and \emph{Civil-Complaint (Call
Centre) Question-Answer Data} --- are outcomes of projects of the National Information Society
Agency (NIA) of the Republic of Korea, distributed through AI-Hub
(\url{https://aihub.or.kr})~\cite{aihubabsa,aihubsmb,aihubcall}. We state this as their terms
require, and we note that neither the corpora nor any excerpt of them is redistributed here.

\section{Conclusion}

We built a system that infers commercial intent without third-party identifiers, turned its
three constraints into falsifiable propositions, and adjudicated them with code. One holds
outright and two hold only on the platforms we could measure: raw tokens undetected at $7/7$; a
deployment payload of $2{,}942{,}652$ bytes ($93.5\%$ of 3\,MiB) on the largest of three measured links, Android; and a Tier-0 $p_{95}$ of $2.00$--$3.37$\,ms on three server CPUs running the same
code, $4.431$ and $3.670$\,ms on two iPhones two SoC generations apart, and $5.080$\,ms on a budget
Android tablet --- $22.2\%$, $18.3\%$ and $25.4\%$ of the 20\,ms budget, with every device slower
than every server CPU. One gate outside that
set is red --- the attribution-axis consistency check --- and \S\ref{sec:gates} says why we
report it rather than wait for it. On the quality axis, the $2.14$\,MB static encoder retains
$78$\,\textasciitilde\,$87\%$ of a $196.6\times$ larger teacher on real-data classification, and
on the product surface it delivers a mid-level (L2) top-5 of $77.6\%$ on colloquial queries in
categories it has never seen, over a $6{,}020$-leaf space.

But what will likely outlast the numbers in this paper are its three negative results.

First, \textbf{turning on the categorical perception channel was a net loss}. A block-level
signal scrambles leaf order within the block, and that damage more than offsets the gain in
block accuracy. The dense-embedding probe quantified that gap at $56.15$\,pp, and that is the
basis on which we closed the channel and pivoted to the rectangular projection.

Second, \textbf{the target is the ceiling}. A projection from images into our space cannot
exceed the target vector. Changing a single thing --- the language, by translating English
titles into Korean --- raised the ceiling by $17.2$\,pp, and the projection rose with it. The
proposition that data is the ceiling holds not only for the training set but for the
\emph{target vector} itself.

Third, \textbf{the moat is not accuracy --- it is access}. If the server received the behavior
signal, it would improve by exactly as much. Our claim is not that we are smarter than the
server, but that we do the same thing even though the signal never leaves the device. And that
claim is adjudicated not by an accuracy metric but by the egress gate and the type system.

What remains is clear: real usage utterances and real labels. The personalization axis stands on
3 real labels; taxonomy supervision carries no deployment-derived Korean labels; and while
\S\ref{sec:aihub} evaluates the deployment path against real Korean commerce text, that text is
a public corpus, not traffic from a device running this SDK. The code is ready, the budget is
ready; what is blocking us is data. So the next step is not a model --- it is a pilot.

\subsection*{Reproducibility}

Every distinctive number in this paper is cited from a ledger file in the repository, and that
claim is checked by code rather than asserted: \code{python3 paper/cite\_audit.py --strict}
extracts every figure carrying three or more decimal places or a thousands separator and fails
unless it can be located in a JSON ledger under \code{experiments/}, \code{artifacts/} or
\code{gates/}. It currently checks $386$ such figures and classifies every one of them; an unclassified figure
fails the run. Nine are derived (sums, differences and ratios of ledger values, each with its
derivation written out), a further group are figures we are not permitted to redistribute or are
quoting in order to retract, and \textbf{exactly one is backed only by a prose specification and
has no ledger at all}. That figure is printed as a defect on every run rather than exempted quietly, because
hiding it is how the problem arose: we built this check after finding that four cited figures came
from a prose note, from a gate that never wrote its result down, or from a measurement taken over
the smaller taxonomy that preceded the expansion. None of the four was caught by reading, and one
of them --- the behavioral gain --- was wrong by $2.8$ percentage points.

Building the check exposed two further defects in the gates themselves. The C3 transmission gate
had been reporting \code{UNMEASURED} because its sample generator fed a $128$-dimensional
embedding into a federated path that consumes only the first $64$; the production path had
followed the encoder from 64 to 128 dimensions and this one generator had not. For as long as that
was true, the paper asserted seven canaries undetected while the gate was measuring nothing. It now
passes, and the measured payload it records is $79{,}968$ bytes rather than the $69{,}482$ an
earlier draft quoted from a prose specification. Separately, nine of the thirteen gates printed
their numbers without writing them anywhere; reporting a verdict and pinning its figures are now
the same function call, so a gate cannot do one without the other. The C9 gate additionally
returned early on an axis mismatch, so the composed $\varepsilon$ the paper cites was computed
nowhere --- the axis check and the composition are independent, and it now completes the
accounting before failing. With those repairs and with \code{dp-accounting} installed, the suite
reports twelve passes, one failure and \emph{no} \code{UNMEASURED} gates against the full working
tree, where at the start of this revision two gates were reporting nothing at all. Gate verdicts were produced by
\code{python3 gates/run\_all.py --strict} run against the full working tree, which holds the
artifacts and crates the published bundle does not; artifact fingerprints come from
\code{artifacts/MANIFEST.json}; the authoritative budget from \code{gates/budgets.json}; and
on-screen figures from \code{artifacts/ui-facts.json}. A reader running the same command against
the bundle alone gets \code{UNMEASURED} wherever an artifact is absent, by design. The final contrastive training stage is
deterministic for a fixed input artifact and seed; the upstream tokenlearn and corpus-preparation
stages are not, and \S\ref{sec:distill} reports the measured drift. Experiment
ledgers are the JSON files in each directory under \code{experiments/}.

The real-data experiments of \S\ref{sec:aihub} are reproducible without our receiving or
redistributing anything. A reader who obtains their own AI-Hub access runs
\code{fetch\_aihub.py} (which downloads the twenty Validation label archives directly, reassembling
the multi-part responses and decoding the CP949 archive entries), then \code{build\_corpus.py},
then \code{aihub\_surface.py}, \code{ir\_baselines.py}, \code{p1\_stats.py},
\code{map\_sensitivity.py} and \code{selective\_risk.py}. All of these run locally, including the
teacher arm: we do not send AI-Hub text to any cluster. Nothing derived from the corpus is
committed --- the repository carries the crosswalk we wrote, the harness code and the aggregate
JSON outputs, and its \code{.gitignore} enforces that. One incidental finding is worth passing on
to anyone scripting AI-Hub downloads: the official shell client's merge routine escapes the shard
prefix with \code{printf \%q} and then passes it to \code{find -name}, which does not match Korean
filenames; the redirect that follows empties the output file and the subsequent \code{rm} deletes
the shards, so the failure presents as a zero-byte archive rather than as an error.


\begin{thebibliography}{99}\small

\bibitem{bm25}
S. Robertson and H. Zaragoza.
\newblock The Probabilistic Relevance Framework: BM25 and Beyond.
\newblock Foundations and Trends in Information Retrieval, 3(4):333--389, 2009.

\bibitem{dpr}
V. Karpukhin, B. O\u{g}uz, S. Min, P. Lewis, L. Wu, S. Edunov, D. Chen, and W. Yih.
\newblock Dense Passage Retrieval for Open-Domain Question Answering.
\newblock EMNLP 2020.

\bibitem{colbert}
O. Khattab and M. Zaharia.
\newblock ColBERT: Efficient and Effective Passage Search via Contextualized Late Interaction
over BERT.
\newblock SIGIR 2020.

\bibitem{pq}
H. J\'egou, M. Douze, and C. Schmid.
\newblock Product Quantization for Nearest Neighbor Search.
\newblock IEEE TPAMI, 33(1):117--128, 2011.

\bibitem{faiss}
J. Johnson, M. Douze, and H. J\'egou.
\newblock Billion-scale similarity search with GPUs.
\newblock IEEE Transactions on Big Data, 2019.

\bibitem{parabel}
Y. Prabhu, A. Kag, S. Gopinath, K. Dahiya, S. Harsola, R. Agrawal, and M. Varma.
\newblock Parabel: Partitioned Label Trees for Extreme Classification with Application to
Dynamic Search Advertising.
\newblock WWW 2018.

\bibitem{kd}
G. Hinton, O. Vinyals, and J. Dean.
\newblock Distilling the Knowledge in a Neural Network.
\newblock arXiv:1503.02531, 2015.

\bibitem{distilbert}
V. Sanh, L. Debut, J. Chaumond, and T. Wolf.
\newblock DistilBERT, a distilled version of BERT: smaller, faster, cheaper and lighter.
\newblock arXiv:1910.01108, 2019.

\bibitem{mobilebert}
Z. Sun, H. Yu, X. Song, R. Liu, Y. Yang, and D. Zhou.
\newblock MobileBERT: a Compact Task-Agnostic BERT for Resource-Limited Devices.
\newblock ACL 2020.

\bibitem{word2vec}
T. Mikolov, I. Sutskever, K. Chen, G. Corrado, and J. Dean.
\newblock Distributed Representations of Words and Phrases and their Compositionality.
\newblock NIPS 2013.

\bibitem{glove}
J. Pennington, R. Socher, and C. D. Manning.
\newblock GloVe: Global Vectors for Word Representation.
\newblock EMNLP 2014.

\bibitem{fasttext}
P. Bojanowski, E. Grave, A. Joulin, and T. Mikolov.
\newblock Enriching Word Vectors with Subword Information.
\newblock TACL, 5:135--146, 2017.

\bibitem{llmint8}
T. Dettmers, M. Lewis, Y. Belkada, and L. Zettlemoyer.
\newblock LLM.int8(): 8-bit Matrix Multiplication for Transformers at Scale.
\newblock NeurIPS 2022.

\bibitem{gptq}
E. Frantar, S. Ashkboos, T. Hoefler, and D. Alistarh.
\newblock GPTQ: Accurate Post-Training Quantization for Generative Pre-trained Transformers.
\newblock ICLR 2023.

\bibitem{infonce}
A. van den Oord, Y. Li, and O. Vinyals.
\newblock Representation Learning with Contrastive Predictive Coding.
\newblock arXiv:1807.03748, 2018.

\bibitem{clip}
A. Radford, J. W. Kim, C. Hallacy, A. Ramesh, G. Goh, S. Agarwal, G. Sastry, A. Askell,
P. Mishkin, J. Clark, G. Krueger, and I. Sutskever.
\newblock Learning Transferable Visual Models From Natural Language Supervision.
\newblock ICML 2021.

\bibitem{flopen}
P. Kairouz, H. B. McMahan, et al.
\newblock Advances and Open Problems in Federated Learning.
\newblock Foundations and Trends in Machine Learning, 14(1--2), 2021.

\bibitem{rappor}
\'U. Erlingsson, V. Pihur, and A. Korolova.
\newblock RAPPOR: Randomized Aggregatable Privacy-Preserving Ordinal Response.
\newblock CCS 2014.

\bibitem{bgnbd}
P. S. Fader, B. G. S. Hardie, and K. L. Lee.
\newblock ``Counting Your Customers'' the Easy Way: An Alternative to the Pareto/NBD Model.
\newblock Marketing Science, 24(2):275--284, 2005.

\bibitem{openrtb}
IAB Technology Laboratory.
\newblock OpenRTB API Specification, version 2.6.
\newblock \url{https://iabtechlab.com/standards/openrtb/}

\bibitem{fedavg}
H.~B. McMahan, E.~Moore, D.~Ramage, S.~Hampson, B.~A. y~Arcas.
\newblock Communication-Efficient Learning of Deep Networks from Decentralized Data.
\newblock \emph{AISTATS}, 2017. arXiv:1602.05629.

\bibitem{fedper}
M.~G. Arivazhagan, V.~Aggarwal, A.~K. Singh, S.~Choudhary.
\newblock Federated Learning with Personalization Layers.
\newblock arXiv:1912.00818, 2019.

\bibitem{dpsgd}
M.~Abadi, A.~Chu, I.~Goodfellow, H.~B. McMahan, I.~Mironov, K.~Talwar, L.~Zhang.
\newblock Deep Learning with Differential Privacy.
\newblock \emph{ACM CCS}, 2016. arXiv:1607.00133.

\bibitem{zcdp}
M.~Bun, T.~Steinke.
\newblock Concentrated Differential Privacy: Simplifications, Extensions, and Lower Bounds.
\newblock \emph{TCC}, 2016. arXiv:1605.02065.

\bibitem{secagg}
K.~Bonawitz et al.
\newblock Practical Secure Aggregation for Privacy-Preserving Machine Learning.
\newblock \emph{ACM CCS}, 2017.

\bibitem{gboard}
Z.~Xu et al.
\newblock Federated Learning of Gboard Language Models with Differential Privacy.
\newblock arXiv:2305.18465, 2023.

\bibitem{sbert}
N.~Reimers, I.~Gurevych.
\newblock Sentence-BERT: Sentence Embeddings using Siamese BERT-Networks.
\newblock \emph{EMNLP-IJCNLP}, 2019. arXiv:1908.10084.

\bibitem{model2vec}
MinishLab.
\newblock Model2Vec: Distill a Small Fast Model from any Sentence Transformer.
\newblock Software. \url{https://github.com/MinishLab/model2vec}

\bibitem{mrl}
A.~Kusupati et al.
\newblock Matryoshka Representation Learning.
\newblock \emph{NeurIPS}, 2022. arXiv:2205.13147.

\bibitem{klue}
S.~Park et al.
\newblock KLUE: Korean Language Understanding Evaluation.
\newblock \emph{NeurIPS Datasets and Benchmarks}, 2021. arXiv:2105.09680.

\bibitem{nsmc}
E.~Park.
\newblock Naver Sentiment Movie Corpus (NSMC).
\newblock Dataset. \url{https://github.com/e9t/nsmc}

\bibitem{esci}
C.~K. Reddy et al.
\newblock Shopping Queries Dataset: A Large-Scale ESCI Benchmark for Improving Product Search.
\newblock arXiv:2206.06588, 2022.

\bibitem{navershop}
bab2min.
\newblock Korean corpus collection, \code{sentiment/naver\_shopping.txt} --- 200,000 Naver
Shopping reviews with ratings. Public domain.
\newblock \url{https://github.com/bab2min/corpus}

\bibitem{aihubabsa}
National Information Society Agency (NIA), Korea.
\newblock Aspect-Based Sentiment Analysis Data (속성기반 감정분석 데이터), dataset 71603.
\newblock AI-Hub, 2021. \url{https://aihub.or.kr}

\bibitem{aihubsmb}
National Information Society Agency (NIA), Korea.
\newblock Small-Business Customer Order Question-Answer Data
(소상공인 고객 주문 질의-응답 데이터), dataset 102.
\newblock AI-Hub, 2021. \url{https://aihub.or.kr}

\bibitem{aihubcall}
National Information Society Agency (NIA), Korea.
\newblock Civil-Complaint (Call Centre) Question-Answer Data
(민원(콜센터) 질의-응답 데이터), dataset 98.
\newblock AI-Hub, 2022. \url{https://aihub.or.kr}

\bibitem{abo}
J.~Collins et al.
\newblock ABO: Dataset and Benchmarks for Real-World 3D Object Understanding.
\newblock \emph{CVPR}, 2022. arXiv:2110.06199.

\bibitem{applevision}
Apple Inc.
\newblock \code{VNGenerateImage\-\allowbreak Feature\-\allowbreak PrintRequest}: Vision framework documentation.
\newblock \url{https://developer.apple.com/documentation/vision/vngenerateimagefeatureprintrequest}

\bibitem{mediapipe}
Google.
\newblock MediaPipe Image Embedder for Android.
\newblock \url{https://ai.google.dev/edge/mediapipe/solutions/vision/image_embedder/android}

\bibitem{mlkit}
Google.
\newblock ML Kit installation paths (bundled vs unbundled).
\newblock \url{https://developers.google.com/ml-kit/tips/installation-paths}

\bibitem{dpacct}
Google.
\newblock \code{dp-accounting}: Privacy Loss Distribution accountant.
\newblock \url{https://github.com/google/differential-privacy}

\bibitem{adnostic}
V. Toubiana, A. Narayanan, D. Boneh, H. Nissenbaum, and S. Barocas.
\newblock Adnostic: Privacy Preserving Targeted Advertising.
\newblock NDSS 2010.

\bibitem{privad}
S. Guha et al.
\newblock Privad: Practical Privacy in Online Advertising.
\newblock NSDI 2011.

\bibitem{fledge}
Google / W3C.
\newblock Protected Audience API (formerly FLEDGE): on-device interest-group bidding.
\newblock \url{https://privacysandbox.google.com/}

\bibitem{chrome2025}
A. Chavez (Google).
\newblock Next steps for Privacy Sandbox and tracking protections in Chrome, 22 April 2025;
and the wind-down of the Privacy Sandbox initiative, October 2025.
\newblock \url{https://privacysandbox.google.com/blog/privacy-sandbox-next-steps}

\bibitem{evidence}
H. Han.
\newblock Evidence bundle: measurement artifacts and verification gates for this paper.
\newblock Commit \code{155b6d94c6ce}.
\newblock \url{https://github.com/hyojunguy/ondevice-intent-evidence}

\bibitem{walmart2026}
A. Kekuda, S. Srirangamsridharan, I. Bhatt, Y. Cao, S. Subramaniam, E. Korpeoglu, K. Nag,
and K. Achan.
\newblock Timing-Aware Repurchase Prediction for Web-Scale E-Commerce: Survival Models for
Multi-Surface Grocery Recommendation.
\newblock arXiv:2608.28393.
\newblock \url{https://arxiv.org/abs/2608.28393}

\bibitem{pipc2026}
Personal Information Protection Commission (Republic of Korea).
\newblock Sanction against TikTok for unlawful collection of behavioural information for
customized advertising, plenary meeting of 22 July 2026 (KRW 10.306\,bn penalty surcharge,
with corrective and disclosure orders).
\newblock Commission announcement, July 2026.
\newblock \url{https://www.pipc.go.kr/}

\end{thebibliography}
\end{document}